# Dynamics of Associative Polymers with High Density of Reversible Bonds

*Shifeng Nian[1, †], Shalin Patil[4, †], Siteng Zhang[5], Myoeum Kim[1], Quan Chen[6], Mikhail Zhernenkov[7], Ting Ge[5], Shiwang Cheng[4,\*], and Li-Heng Cai* (蔡历恒)[1,2,3,\*]

[1]Soft Biomatter Laboratory, Department of Materials Science and Engineering, University of Virginia, Charlottesville, VA 22904, USA
[2]Department of Chemical Engineering, University of Virginia, Charlottesville, VA 22904, USA
[3]Department of Biomedical Engineering, University of Virginia, Charlottesville, VA 22904, USA
[4]Department of Chemical Engineering and Materials Science, Michigan State University, East Lansing, MI 48824, USA
[5]Department of Chemistry and Biochemistry, University of South Carolina, Columbia, SC 29208, USA
[6]State Key Lab Polymer Physics and Chemistry, Changchun Institute of Applied Chemistry, Renmin St. 5625, Changchun 130022, Jilin, P. R. China
[7]National Synchrotron Light Source-II, Brookhaven National Laboratory, Upton, NY 11973, USA

[\*]Corresponding authors: L.-H. C., liheng.cai@virginia.edu; S. C., chengsh9@msu.edu.

[†]Equal contribution.

**Corresponding author contact:**
  Dr. Li-Heng Cai
  228 Wilsdorf Hall
  University of Virginia
  395 McCormick Road
  Charlottesville, VA 22904
  Tel: 434-924-2512
  Fax: 434-982-5660

**Word Count: 3747**
  Abstract: 129
  Text word count: 2949
  Equations word equiv.: 64
  Figures word equiv.: 734

**Abstract:** We design and synthesize unentangled associative polymers carrying unprecedented high fractions of stickers, up to eight per Kuhn segment, that can form strong pairwise hydrogen bonding of ~$20k_BT$ without microphase separation. The reversible bonds significantly slow down the polymer dynamics but nearly do not change the shape of linear viscoelastic spectra. Moreover, the structural relaxation time of associative polymers increases exponentially with the fraction of stickers and exhibits a universal yet non-Arrhenius dependence on the distance from polymer glass transition temperature. These results cannot be understood within the framework of the classic sticky-Rouse model but are rationalized by a renormalized Rouse model, which highlights an unexpected influence of reversible bonds on the structural relaxation rather than the shape of viscoelastic spectra for associative polymers with high concentrations of stickers.

**Significance Statement**

An associative polymer carries many stickers that can form reversible associations. For > 30 years, the understanding is that reversible associations change the shape of linear viscoelastic spectra by adding a plateau in the intermediate frequency range, at which associations have not yet relaxed and thus effectively act as crosslinks. We experimentally show that this molecular picture is incorrect for homogenous associative polymers carrying high density of stickers. Our experimental discovery is rationalized by a renormalized Rouse model that highlights an unexpected influence of reversible interactions on the structural relaxation rather than the shape of viscoelastic spectra of associative polymers.

Main text

An associative polymer carries many moieties that can form reversible bonds (ref. [1] and therein). Unlike permanent covalent bonds, a reversible association can break and reform at laboratory time scales [2–9]. This process not only slows down polymer dynamics but also dissipates energy, enabling macroscopic properties inaccessible by conventional polymers. As a result, associative polymers are widely used as viscosity modifiers for fuels [10], lubricants [11], and paints [12], to create tough self-healing polymers [13,14] and reprocessable supramolecular polymer networks [15–19], and to engineer biomaterials with prescribed dynamic mechanical properties critical to tissue engineering and regeneration [20,21]. Thus, understanding the effects of reversible interactions on the dynamics of associative polymers is of both technological and fundamental importance.

All existing understanding of associative polymers is built on a fundamental timescale – the *lifetime of a reversible association* $\tau_s$, which increases exponentially with the activation energy $E_a$:

$$\tau_s = \tau_0 \exp\left(\frac{E_a}{k_B T}\right) \quad (1)$$

Here, $k_B$ is Boltzmann constant, $T$ is absolute temperature, and $\tau_0$ the relaxation time of a Kuhn monomer in the absence of any attraction between stickers. Equation (1) implies two widely accepted physical consequences: (i) the structural relaxation time $\tau_0$ is independent of the concentration of stickers; (ii) introducing associations changes the viscoelastic spectra by widening the separation between the structural relaxation and the terminal flow, and the width of the separation increases exponentially with the activation energy.

The activation energy, $E_a$, is defined as the strength of an association. For relatively simple pairwise associations such as hydrogen bonding, the activation energy has been assumed to be a constant determined by the bond strength. This is the *foundational assumption* for all existing theoretical models (refs. [22–26] and therein) including ours [27]. Experimentally, the activation energy is measured through the change of the dynamics of associative polymers [28–31]. However, in *almost all* existing experimental systems, the strong interaction between stickers often leads to nanoscale aggregations or even microphase separation; examples include clusters formed by hydrogen-bonding groups at the two ends of a linear telechelic polymer [10,25,30,32–36] and π-π stacking of quadruple hydrogen bonding formed by 2-ureido-4[1H]-pyrimidinone (UPy) groups [37]. Because the nanoscale cluster is distinct from its surrounding environment, it creates both entropic and enthalpic barriers that prevent dissociation; this precludes a precise interpretation of experimentally measured activation energy. As a result, it has yet to be rigorously tested the relation between activation energy and bond strength.

In this *Letter*, we seek to experimentally answer two fundamental questions about associative polymers. *First*, what is the relation between the activation energy and the bond strength of a pairwise association? *Second*, would the reversible interactions affect the shape of linear viscoelastic spectra of associative polymers? We are interested in polymers with high concentrations of stickers close to and higher than one per Kuhn segment. Such high concentrations often occur in experimental systems, and sometimes are even required to create polymeric materials with optimized properties such as high stiffness and rapid self-healing ability [17,18,38]. Yet little is understood about the dynamics of associative polymers with high density of reversible bonds.

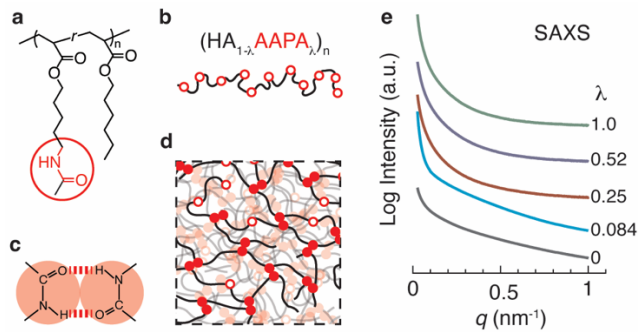

**Figure 1. Molecular design and structure of associative polymers. (a)** An associative polymer is synthesized by copolymerizing hexyl acrylate (HA) and 5-acetamido-1-pentyl acrylate (AAPA). **(b)** Open circles: unpaired amide groups ("open" stickers); $\lambda$: fraction of stickers; $n$: degree of polymerization (DP). **(c)** Two amide groups form a pairwise hydrogen bond of strength $\epsilon_b \approx 20 k_B T$. Solid circles: "closed" stickers. **(d)** A schematic of an associative polymer network. **(e)** Small-angle X-ray scattering (SAXS) intensity profiles of all associative polymers.

We design model associative polymers using amide groups as stickers, which form pairwise double hydrogen bonding without inducing microphase separation at high concentrations of stickers, as experimentally demonstrated in our previous work [38]. Yet, in Ref. [38] the polymer is a randomly branched molecule lacking control over molecular architecture and molecular weight (MW), preventing quantitative understanding of polymer dynamics. In the current study, we copolymerize two monomers, hexyl acrylate (HA) and 5-acetamido-1-pentyl acrylate (AAPA), to create an associative polymer, $(HA_{1-\lambda}AAPA_\lambda)_n$, in which $n$ is the degree of polymerization (DP) and $\lambda$ is the fraction of the sticky monomer AAPA (**Fig. 1a&b**). Two sticky monomers can form an amide-amide double hydrogen bond (a pair of closed circles in **Fig. 1c**) that crosslinks the polymers to form a transient network (**Fig. 1d**). Compared to the spacer monomer HA, the sticky monomer AAPA is essentially the same except that it carries an amide group at one of its ends (**Fig. 1a**). Thus, it is reasonable to assume that the Kuhn length, $b$, of a reversible polymer does not change with $\lambda$. Each Kuhn segment consists of on average 8.7 chemical repeating units and that $b = 22$ Å (see **SI Text**). The calculated entanglement modulus of PHA, $G_e = 110$ kPa, agrees well with experiments (**Fig. S2**). These results provide the basic polymer physics parameters of associative polymers.

We develop a procedure to synthesize a series of associative polymers with fixed DP ~250 but various $\lambda$ from 0, 0.084, 0.25, 0.52, to 1.0 (**Fig. S1**, **Tables S1&2**); these correspond to on average 0, 0.7, 2, 4, and 8 stickers per Kuhn segment. Moreover, the MW of the polymers, 39 kDa, is below the critical MW $M_c \approx 46$ kDa, such that the effects of entanglements are negligible (see **SI Text**). All polymers form homogenous, amorphous liquids without nanoscale clusters regardless of the fraction of stickers, as evidenced by the absence of characteristic peaks at the low and intermediate wavevectors, $q$, from small-angle X-ray scattering (SAXS) measurements (**Fig. 1e**). Importantly, the amide-amide bonding strength relatively strong, $\epsilon_b \approx 20k_BT$ [39], such that their effects on polymer dynamics, if any, can be experimentally revealed. Thus, our polymers provide an ideal model system to explore the effects of pure pairwise bonding on the behavior of associative polymers.

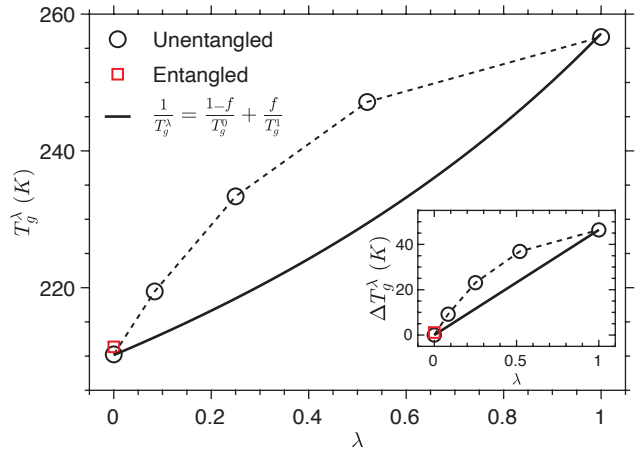

**Figure 2.** The glass transition temperature $T_g^\lambda$ of associative polymers with $\lambda$ stickers. Solid line: Fox relation for $T_g^\lambda$ using the weight fraction $f$ of stickers as the variable. Circles: unentangled polymers with DP ~250; square: an entangled polymer with DP 1009. Dashed line is the guidance for the eye. Inset: $\Delta T_g^\lambda \equiv T_g^\lambda - T_g^0$ vs $\lambda$. Solid line: linear dependence predicted by a recent theory for telechelic associative polymers [25].

The addition of stickers dramatically increases the glass transition temperature, $T_g$, of associative polymers, as shown by differential scanning calorimetry measurements in **Fig. S3**. As $\lambda$ increases from 0 to 1, $T_g$ increases by nearly 50°C from -64.5°C to -16.5°C (circles in **Fig. 2**). Yet, increasing the polymer MW by four times from 39 to 157 kDa does not alter the $T_g$ (square in **Fig. 2**). Interestingly, the measured $T_g$ is higher than that predicted by the Fox relation [40]: $1/T_g^\lambda = \Sigma_i f_i/T_g^i$, where $T_g^i$ is the $T_g$ of homopolymer $i$, and $f_i$ is the mass fraction of monomer units $i$ in the copolymer (solid line in **Fig. 2**). Moreover, the change of $T_g$ exhibits a nonlinear, convex-like dependence on the fraction of stickers, contrasting the linear dependence for telechelic associative polymers with nanoscale aggregations [25] (inset of **Fig. 2**). These differences highlight the unique behavior of associative polymers carrying high concentrations of stickers without microphase separation. Importantly, the dramatic increase in $T_g$ implies that reversible interactions enhance monomeric friction to slow down polymer dynamics.

To further explore the effects of reversible associations on polymer dynamics, we use a stress-controlled rheometer equipped with an environmental chamber to quantify the linear viscoelasticity of polymers (see **SI Materials and Methods**). We use time-temperature superposition [41] to construct master curves for the dependencies of storage $G'(\omega)$ loss $G''(\omega)$ moduli on oscillatory shear frequency $\omega$. Remarkably, the dynamics of the polymer with on average 0.7 stickers per Kuhn segment ($\lambda = 0.084$) is nearly identical to that of the control polymer without stickers (**Fig. 3**). By contrast, existing theories predict that, for a bonding strength $\epsilon_b \approx 20\, k_B T$, the terminal relaxation should be slowed by at least $\tau_s/\tau_0 \sim \exp(\epsilon_b/k_B T) \sim 10^8$ times [eq. (1)]. Such a dramatic discrepancy unambiguously demonstrates an unexpected weak influence of strong reversible bonds on the linear viscoelasticity of associative polymers without microphase separation.

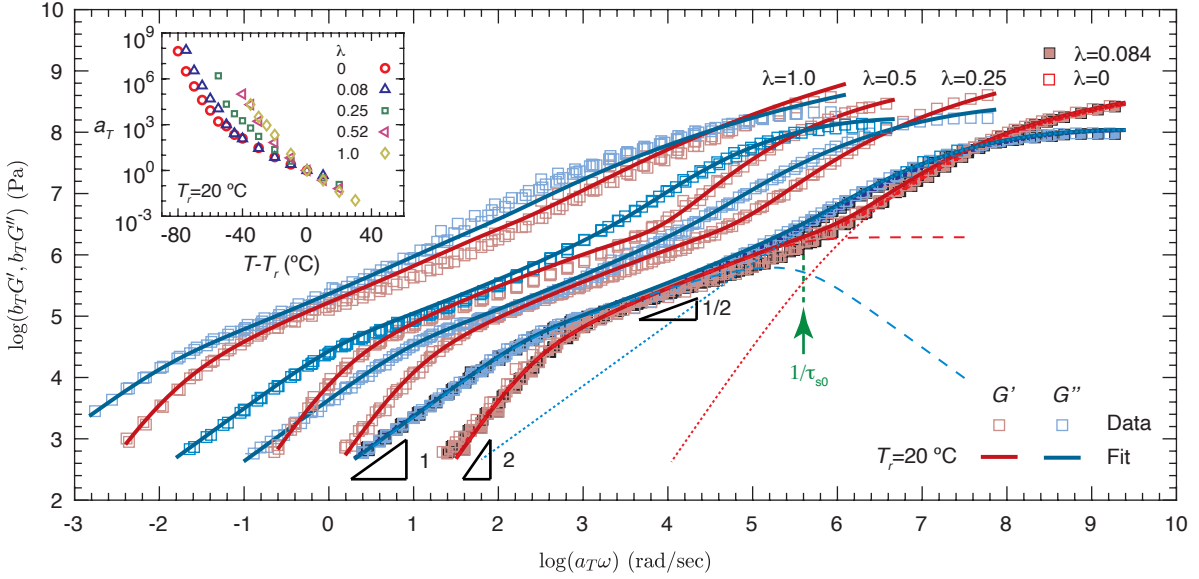

**Figure 3. Linear viscoelasticity of unentangled associative polymers.** Master curves of storage ($G'$) and loss ($G''$) moduli as a function of angular frequency $\omega$ for unentangled polymers at a reference temperature of $T_r = 20°C$. Solid lines represent the fit of the sum of *renormalized* Rouse model for low-frequency behavior, $G_{rR}(t)$ [eq. (2)], and Kohlrausch–Williams–Watts (KWW) model [42], $G_g(t) = G_g(0) \exp\left[-\left(\frac{t}{\tau_{KWW}}\right)^\beta\right]$, for high-frequency glassy dynamics: $G(t) = G_{rR}(t) + G_g(t)$ (see **SI Text**). Here, $G_g(0)$ is the glassy modulus at $t \to 0$, $\tau_{KWW}$ is the characteristic time of the glassy relaxation, the exponent $\beta$ describes the distribution of glassy relaxation modes: The lower the value of $\beta$, the broader the mode distribution (see **SI Text**). All the fitting parameters are listed in **Table S1**. Dotted lines are an example of using KWW model to fit the glassy dynamics of polymers, and dashed lines are an example of using the *renormalized* Rouse model to fit the low-frequency relaxation modulus. The modulus shift factor $b_T = \rho T/\rho_r T_r$, where $\rho_r$ is the polymer density at the reference temperature $T_r$, and $T$ is the absolute temperature at which the oscillatory frequency measurements are performed. Inset: time scale shift factor $a_T$ for all associative polymers.

Further increasing the fraction of stickers significantly shifts the master curves to lower frequencies (**Fig. 3**). Yet, the master curves can be horizontally shifted to nearly perfect overlap with each other up to intermediate frequency (**Fig. S4**). These results show that reversible associations slow down the polymer dynamics but almost do not change the shape of the relaxation spectra regardless of the concentration of stickers. The observed behavior is in striking contrast to that predicted by the classic sticky-Rouse model for unentangled associative polymers: reversible associations change the shape of linear viscoelastic spectra by adding a plateau in the intermediate frequency range, at which associations have not yet relaxed and thus effectively act as crosslinks [30,32,34,43–46]. Collectively, our experimental results demonstrate that the classic sticky-Rouse model does not apply to associative polymers at sticker concentrations higher than one per Kuhn segment.

To explain the remarkable dynamics of associative polymers carrying high concentrations of stickers, we propose a *renormalized* Rouse model. Instead of considering a reversible polymer as a copolymer consisting of the spacer and sticky monomers, we treat it as a homopolymer consisting of *renormalized* Kuhn segments with a relaxation time $\tau_{s0}$. Thus, the Rouse relaxation modulus for polymers with $\lambda$ stickers is:

$$G_{rR}(t,\lambda) = \sum_i N_{Av} k_B T \frac{\rho w_{i,\lambda}}{M_{i,\lambda}} \sum_{q=1}^{N_{i,\lambda}} \exp\left(-\frac{tq^2}{\tau_{s0} N_{i,\lambda}^2}\right) \qquad (2)$$

in which $w_{i,\lambda}$, $M_{i,\lambda}$, and $N_{i,\lambda} = M_{i,\lambda}/M_0(\lambda)$ are, respectively, the weight fraction, MW, and the number of Kuhn monomers of the $i$-th polymer with $\lambda$ stickers. Here $M_0(\lambda) = n_k[m_{HA}(1-\lambda) + m_{AAPA}\lambda]$ is the mass of a Kuhn monomer with $\lambda$ stickers, in which $m_{HA} = 156$ Da is the mass of the spacer monomer HA, $m_{AAPA} = 199$ Da is the mass of the sticky monomer AAPA, and the $n_k = 8.7$ is the number of chemical monomers per Kuhn segment.

We emphasize that $\tau_{s0}$ cannot be viewed as the renormalized lifetime of a sticker in the classic sticky-Rouse model [27,36]. In this model, the renormalized lifetime is defined as the average time from the first formation of bond between a particular pair of stickers until the formation of a bond with a new open sticker. Thus, it becomes easier for an open sticker to find a new partner at a higher concentration of stickers. And the sticky-Rouse model predicts that the renormalized lifetime of a sticker decreases with the increase of sticker concentration by a power law, $\sim N_s^{1/6}$, in which $N_s$ is the number of the Kuhn segments between two neighboring stickers [27]. By contrast, for the current associative polymers, each *renormalized* Kuhn segment is an *effective* sticker, and the concentration of *effective* stickers is always a constant one. Thus, $\tau_{s0}$ can be viewed as the basic monomeric relaxation time of an effective sticker.

We quantify $\tau_{s0}$ by using eq. (2) to fit the low $\omega$ relaxation behavior of the measured moduli following a previously described procedure [28]. In eq. (2), the MW distribution $w_{i,\lambda}$ is directly determined from size exclusion chromatography (see **SI Materials and Methods**), and the basic polymer physics parameters are pre-determined from synthesis. Thus, all input parameters for the *renormalized* Rouse model are given by experiments except $\tau_{s0}$. Using $\tau_{s0}$ as the only adjustable parameter, we obtain nearly perfect fit of the model to experiments (lines in **Fig. 3B**). This one-parameter fitting allows precise quantification of $\tau_{s0}(\lambda, T_r)$ for each associative polymer at the reference temperature (symbols in **Fig. 4a**, **Table S1**).

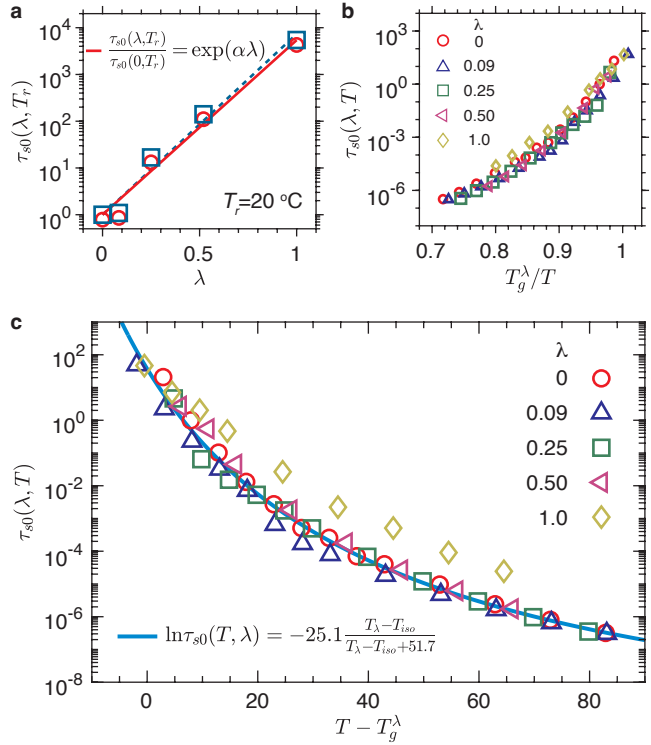

**Figure 4. (a)** Relaxation time of a renormalized Kuhn monomer, $\tau_{s0}$, at $T_r =20$ °C. Lines are the best fit to experiments using eq. (3). Dashed line is for data (squares) using the relaxation of a pure PHA monomer as $\tau_{s0}(0,T_r)$, and the fitting parameter $\alpha = 8.9 \pm 1.2$ [eq. (3)]. Solid line is for data (circles) with the correction of $\tau_{s0}$ due to the slight increase in monomer mass with $\lambda$, and the fitting parameter $\alpha = 8.6 \pm 1.2$. **(b)** The segmental relaxation time $\tau_{s0}(\lambda, T)$ at various temperatures and fractions of stickers collapse to a master curve against reduced temperature $T_g^\lambda/T$. **(c)** $\tau_{s0}$ exhibits a non-Arhuenius dependence of the distance from polymer glass transition temperature. Solid line: the best fit to all data points using eq. (4).

Interestingly, at room temperature the segmental relaxation time increases nearly exponentially with $\lambda$:

$$\tau_{s0}(\lambda, T_r) = \tau_{s0}(0, T_r) \exp(\alpha \lambda) \qquad (3)$$

in which $\tau_{s0}(0, T_r) = 3.7 \times 10^{-7}$ sec is the relaxation time of a PHA Kuhn monomer without reversible interactions (see **Table S1**), and $\alpha$ is a fitting parameter of 8.9±1.2 (dashed line to squares in **Fig. 4a**). Accounting for the correction of $\tau_{s0}$ due to the slight increase in monomer mass with $\lambda$ results in a negligible difference, as confirmed by the nearly the same value $\alpha$=8.6±1.2 (solid line to circles in **Fig. 4a**). Expression (3) indicates that at room temperature the apparent activation energy increases linearly with the concentration of stickers, $E_a/k_B T_r = \alpha\lambda$. This behavior contrasts the widely accepted understanding that the activation energy is independent of the concentration of stickers [22–27]. Yet, our experimental observation can be qualitatively captured by atomistic simulation for melts of pure Kuhn monomers at high temperatures, which reveals the relaxation time of a Kuhn monomer increases exponentially with the fraction of stickers (see **SI Text** and **Fig. S5**). These results show no direct connection between the strength of a pairwise bond and the absolute value of the activation energy.

To explore the dependence of segmental relaxation time on temperature, we construct the master curves at various reference temperatures, and use the same procedure to determine $\tau_{s0}(\lambda, T)$. Remarkably, plotting $\tau_{s0}(\lambda, T)$ against $T_g^\lambda/T$ collapses almost all monomeric relaxation times to a *universal* curve (**Fig. 4b**). Similarly, a universal behavior is observed by plotting $\tau_{s0}(\lambda, T)$ against the distance from polymer glass transition temperature, $T - T_g^\lambda$ (symbols in **Fig. 4c**). Moreover, such a universal behavior can be described by a non-Arrhenius dependence on temperature (solid line in **Fig. 4c**):

$$\ln \tau_{s0}(T, \lambda) = -25.1 \frac{T_\lambda - T_{iso}}{T_\lambda - T_{iso} + 51.7} \tag{4}$$

in which $T_\lambda \equiv T - T_g^\lambda$, and $T_{iso} = 6.5°C$. Equation (4) indicates that, for any fixed fraction of stickers, the temperature dependence of the ratio between the lifetime of a sticky Kuhn segment to that of a Kuhn segment without stickers, $\frac{d}{d(1/T)} \ln \left[\frac{\tau_{s0}(T, \lambda \neq 0)}{\tau_{s0}(T, 0)}\right]$, is not a constant. Thus, for associative polymers with a given $\lambda$, a universal activation energy cannot be obtained across the whole temperature range of the measurements. This behavior is fundamentally different from the classical understanding of associative polymers that the ratio exhibits an Arrhenius dependence on temperature, which yields a constant activation energy universal to all temperatures [eq. (1)]. Understanding the molecular origin of the effects of reversible interactions on structural relaxation is beyond the scope of this work and will be the subject of future explorations. Nevertheless, our results show that the *renormalized* Rouse model provides a universal description of the dynamics of associative polymers with high densities of reversible bonds: Reversible interactions slow down the polymer dynamics by decreasing the elementary time scale associated with structural relaxation without changing the shape of linear viscoelastic spectra of polymers.

In summary, we have developed unentangled associative polymers carrying unprecedented high concentrations of stickers that can form pairwise interactions. Unlike conventional associative polymers in which stickers aggregate to nanoscale domains, our polymers form a homogenous network without microphase separation regardless of the concentrations of stickers. The reversible interactions significantly slow down the polymer dynamics but nearly do not change the shape of linear viscoelastic spectra. This behavior contradicts that of conventional associative polymers following the classic sticky-Rouse dynamics. Yet, our experimentally measured linear viscoelasticity can be well described by a *renormalized* Rouse model, in which a reversible polymer is treated as a homopolymer consisting of *renormalized* Kuhn monomers. Remarkably, the segmental relaxation time increases exponentially with the concentration of stickers and exhibits a universal yet non-Arrhenius dependence on the distance from polymer glass transition temperature. We note that reversible interactions widen the width of the glassy transition zone (**Fig. 3**, **Table S1**, and **SI Text**). This behavior is likely because reversible interactions promote local alignment of Kuhn segments to facilitate their cooperative motion; this would increase the number of relaxation modes and thus broaden the glassy transition zone. Further understanding of the glassy dynamics of associative polymers is beyond the scope of this work and will be the subject of future exploration. Nevertheless, our findings reveal an unexpected influence of reversible interactions on the structural relaxation of associative polymers rather than their viscoelastic properties. Thus, our associative polymers provide a system that allows for investigating separately the effects of reversible interactions on chain relaxation and glassy dynamics; this may offer opportunities to improve the understanding of the challenging physics of glass relaxation of polymers [47]. Perhaps more importantly, our discoveries show that the fraction of stickers, not the conventionally thought sticker-sticker binding energy, is a dominant parameter controlling the dissociation dynamics of associative polymers without phase separation, and thus represent a paradigm shift in the development of supramolecular materials [17,20].

**Supporting Information:** Supporting text on chemical synthesis, Kuhn length, glass transition temperature, glassy dynamics, entanglement modulus of associative polymers, and atomistic simulations. Material and methods for polymer synthesis and characterization. $^1$H NMR spectra of all associative polymers.

**Acknowledgements:** L.H.C. acknowledges the support from NSF (CAREER DMR-1944625), ACS Petroleum Research Fund (PRF) (6132047-DNI), Virginia Commonwealth Health Research Grant, and Juvenile Diabetes Research Foundation.

**Author contributions:** L.H.C. conceived and oversaw the study. L.H.C., S.N., S.C, and S.P. designed the research. S.N. synthesized and characterized the polymers and M.K. helped with the synthesis. S.P. performed DSC and rheological measurements under S.C.'s supervision. S.Z. performed atomistic simulation under T.G.'s supervision. Q.C. provided initial code for fitting the linear viscoelasticity data. M.Z. helped with SAXS measurements. L.H.C. and S.C. analyzed the data. L.H.C. developed theoretical models and wrote the paper. All authors reviewed and commented on the paper.

**Competing interests:** L.H.C. and S.N. have filed a patent for the synthesis reversible polymers.

**References**

[1] Z. Zhang, Q. Chen, and R. H. Colby, *Dynamics of Associative Polymers*, Soft Matter **14**, 2961 (2018).

[2] L. Leibler, M. Rubinstein, and R. H. Colby, *Dynamics of Reversible Networks*, Macromolecules **24**, 4701 (1991).

[3] L. G. Baxandall, *Dynamics of Reversibly Crosslinked Chains*, Macromolecules **22**, 1982 (1989).

[4] M. Rubinstein and A. N. Semenov, *Thermoreversible Gelation in Solutions of Associating Polymers. 2. Linear Dynamics*, Macromolecules **31**, 1386 (1998).

[5] M. E. Cates, *Reptation of Living Polymers: Dynamics of Entangled Polymers in the Presence of Reversible Chain-Scission Reactions*, Macromolecules **20**, 2289 (1987).

[6] M. Rubinstein and A. N. Semenov, *Dynamics of Entangled Solutions of Associating Polymers*, Macromolecules **34**, 1058 (2001).

[7] A. N. Semenov and M. Rubinstein, *Dynamics of Entangled Associating Polymers with Large Aggregates*, Macromolecules **35**, 4821 (2002).

[8] A. N. Semenov, J. F. Joanny, and A. R. Khokhlov, *Associating Polymers: Equilibrium and Linear Viscoelasticity*, Macromolecules **28**, 1066 (1995).

[9] F. Tanaka and S. F. Edwards, *Viscoelastic Properties of Physically Cross-Linked Networks. Transient Network Theory*, Macromolecules **25**, 1516 (1992).

[10] M.-H. Wei, B. Li, R. L. A. David, S. C. Jones, V. Sarohia, J. A. Schmitigal, and J. A. Kornfield, *Mega-Supramolecules for Safer, Cleaner Fuel by End-Association of Long Telechelic Polymers*, Science **350**, 72 (2015).

[11] T. J. Murdoch, E. Pashkovski, R. Patterson, R. W. Carpick, and D. Lee, *Sticky but Slick: Reducing Friction Using Associative and Nonassociative Polymer Lubricant Additives*, ACS Appl Polym Mater **2**, 4062 (2020).

[12] A. J. Reuvers, *Control of Rheology of Water-Borne Paints Using Associative Thickeners*, Prog Org Coatings **35**, 171 (1999).

[13] S. C. Grindy, R. Learsch, D. Mozhdehi, J. Cheng, D. G. Barrett, Z. Guan, P. B. Messersmith, and N. Holten-Andersen, *Control of Hierarchical Polymer Mechanics with Bioinspired Metal-Coordination Dynamics*, Nat Mater **14**, 1210 (2015).

[14] E. Filippidi, T. R. Cristiani, C. D. Eisenbach, J. Herbert Waite, J. N. Israelachvili, B. Kollbe Ahn, M. T. Valentine, J. H. Waite, J. N. Israelachvili, B. Kollbe Ahn, and M. T. Valentine, *Toughening Elastomers Using Mussel-Inspired Iron-Catechol Complexes*, Science **358**, 502 (2017).

[15] B. J. B. Folmer, R. P. Sijbesma, R. M. Versteegen, J. a J. Van Der Rijt, and E. W. Meijer, *Supramolecular Polymer Materials: Chain Extension of Telechelic Polymers Using a Reactive Hydrogen-Bonding Synthon*, Adv Mater **12**, 874 (2000).

[16] G. R. Gossweiler, G. B. Hewage, G. Soriano, Q. Wang, G. W. Welshofer, X. Zhao, and S. L. Craig, *Mechanochemical Activation of Covalent Bonds in Polymers with Full and Repeatable Macroscopic Shape Recovery*, ACS Macro Lett **3**, 216 (2014).

[17] P. Cordier, F. Tournilhac, C. Soulié-Ziakovic, L. Leibler, C. Soulie-Ziakovic, and L. Leibler, *Self-Healing and Thermoreversible Rubber from Supramolecular Assembly*, Nature **451**, 977 (2008).

[18] Y. L. Chen, A. M. Kushner, G. A. Williams, and Z. B. Guan, *Multiphase Design of Autonomic Self-Healing Thermoplastic Elastomers*, Nat Chem **4**, 467 (2012).

[19] M. Burnworth, L. Tang, J. R. Kumpfer, A. J. Duncan, F. L. Beyer, G. L. Fiore, S. J. Rowan,

[20] M. J. Webber, E. A. Appel, E. W. Meijer, and R. Langer, *Supramolecular Biomaterials*, Nat Mater **15**, 13 (2015).

[21] A. M. Rosales and K. S. Anseth, *The Design of Reversible Hydrogels to Capture Extracellular Matrix Dynamics*, Nat Rev Mater **1**, 1 (2016).

[22] D. Amin, A. E. Likhtman, and Z. Wang, *Dynamics in Supramolecular Polymer Networks Formed by Associating Telechelic Chains*, Macromolecules **49**, 7510 (2016).

[23] R. S. Hoy and G. H. Fredrickson, *Thermoreversible Associating Polymer Networks. I. Interplay of Thermodynamics, Chemical Kinetics, and Polymer Physics*, J Chem Phys **131**, 1 (2009).

[24] A. Ghosh and K. S. Schweizer, *Physical Bond Breaking in Associating Copolymer Liquids*, ACS Macro Lett **10**, 122 (2021).

[25] A. Ghosh, S. Samanta, S. Ge, A. P. Sokolov, and K. S. Schweizer, *Influence of Attractive Functional Groups on the Segmental Dynamics and Glass Transition in Associating Polymers*, Macromolecules **55**, 2345 (2022).

[26] A. Ghosh and K. S. Schweizer, *Microscopic Theory of the Effect of Caging and Physical Bonding on Segmental Relaxation in Associating Copolymer Liquids*, Macromolecules **53**, 4366 (2020).

[27] E. B. Stukalin, L.-H. Cai, N. A. Kumar, L. Leibler, and M. Rubinstein, *Self-Healing of Unentangled Polymer Networks with Reversible Bonds*, Macromolecules **46**, 7525 (2013).

[28] Q. Chen, G. J. Tudryn, and R. H. Colby, *Ionomer Dynamics and the Sticky Rouse Model*, J Rheol **57**, 1441 (2013).

[29] Z. Zhang, C. Huang, R. A. Weiss, and Q. Chen, *Association Energy in Strongly Associative Polymers*, J Rheol **61**, 1199 (2017).

[30] A. Mordvinkin, D. Döhler, W. H. Binder, R. H. Colby, and K. Saalwächter, *Terminal Flow of Cluster-Forming Supramolecular Polymer Networks: Single-Chain Relaxation or Micelle Reorganization?*, Phys Rev Lett **125**, 127801 (2020).

[31] S. Tang, M. Wang, and B. D. Olsen, *Anomalous Self-Diffusion and Sticky Rouse Dynamics in Associative Protein Hydrogels*, J Am Chem Soc **137**, 3946 (2015).

[32] T. Yan, K. Schröter, F. Herbst, W. H. Binder, and T. Thurn-Albrecht, *Nanostructure and Rheology of Hydrogen-Bonding Telechelic Polymers in the Melt: From Micellar Liquids and Solids to Supramolecular Gels*, Macromolecules **47**, 2122 (2014).

[33] K. Xing, M. Tress, P. Cao, S. Cheng, T. Saito, V. N. Novikov, and A. P. Sokolov, *Hydrogen-Bond Strength Changes Network Dynamics in Associating Telechelic PDMS*, Soft Matter **14**, 1235 (2018).

[34] S. Ge, S. Samanta, B. Li, G. P. Carden, P. F. Cao, and A. P. Sokolov, *Unravelling the Mechanism of Viscoelasticity in Polymers with Phase-Separated Dynamic Bonds*, ACS Nano **16**, 4746 (2022).

[35] S. Ge, S. Samanta, M. Tress, B. Li, K. Xing, P. Dieudonné-George, A. C. Genix, P. F. Cao, M. Dadmun, and A. P. Sokolov, *Critical Role of the Interfacial Layer in Associating Polymers with Microphase Separation*, Macromolecules **54**, 4246 (2021).

[36] S. Ge, M. Tress, K. Xing, P. F. Cao, T. Saito, and A. P. Sokolov, *Viscoelasticity in Associating Oligomers and Polymers: Experimental Test of the Bond Lifetime Renormalization Model*, Soft Matter **16**, 390 (2020).

[37] D. J. M. Van Beek, A. J. H. Spiering, G. W. M. Peters, K. Te Nijenhuis, and R. P. Sijbesma, *Unidirectional Dimerization and Stacking of Ureidopyrimidinone End Groups in*

[37]    *Polycaprolactone Supramolecular Polymers*, Macromolecules **40**, 8464 (2007).

[38]    J. Wu, L.-H. Cai, and D. A. D. A. Weitz, *Tough Self-Healing Elastomers by Molecular Enforced Integration of Covalent and Reversible Networks*, Adv Mater **29**, 1702616 (2017).

[39]    A. J. Doig and D. H. Williams, *Binding-Energy of an Amide-Amide Hydrogen Bond in Aqueous and Nonpolar Solvents*, J Am Chem Soc **114**, 338 (1992).

[40]    P. Bonardelli, G. Moggi, and A. Turturro, *Glass Transition Temperatures of Copolymer and Terpolymer Fluoroelastomers*, Polymer **27**, 905 (1986).

[41]    M. Rubinstein and R. H. Colby, *Polymer Physics* (Oxford University Press, Oxford, UK, 2003).

[42]    R. Böhmer, K. L. Ngai, C. A. Angell, and D. J. Plazek, *Nonexponential Relaxations in Strong and Fragile Glass Formers*, J Chem Phys **99**, 4201 (1993).

[43]    T. Yan, K. Schröter, F. Herbst, W. H. Binder, and T. Thurn-Albrecht, *What Controls the Structure and the Linear and Nonlinear Rheological Properties of Dense, Dynamic Supramolecular Polymer Networks?*, Macromolecules **50**, 2973 (2017).

[44]    S. Wu and Q. Chen, *Advances and New Opportunities in the Rheology of Physically and Chemically Reversible Polymers*, Macromolecules **55**, 697 (2022).

[45]    X. Cao, X. Yu, J. Qin, and Q. Chen, *Reversible Gelation of Entangled Ionomers*, Macromolecules **52**, 8771 (2019).

[46]    Q. Chen, C. Huang, R. A. Weiss, and R. H. Colby, *Viscoelasticity of Reversible Gelation for Ionomers*, Macromolecules **48**, 1221 (2015).

[47]    B. Mei, Y. Zhou, and K. S. Schweizer, *Experimental Test of a Predicted Dynamics-Structure-Thermodynamics Connection in Molecularly Complex Glass-Forming Liquids*, Proc Natl Acad Sci U S A **118**, 1 (2021).

# Dynamics of Associative Polymers with High Density of Reversible Bonds

*Shifeng Nian[1, †], Shalin Patil[4, †], Siteng Zhang[5], Myoeum Kim[1], Quan Chen[6], Mikhail Zhernenkov[7], Ting Ge[5], Shiwang Cheng[4,\*], and Li-Heng Cai[1,2,3,\*]*

[1]Soft Biomatter Laboratory, Department of Materials Science and Engineering, University of Virginia, Charlottesville, VA 22904, USA
[2]Department of Chemical Engineering, University of Virginia, Charlottesville, VA 22904, USA
[3]Department of Biomedical Engineering, University of Virginia, Charlottesville, VA 22904, USA
[4]Department of Chemical Engineering and Materials Science, Michigan State University, East Lansing, MI 48824, USA
[5]Department of Chemistry and Biochemistry, University of South Carolina, Columbia, SC 29208, USA
[6]State Key Lab Polymer Physics and Chemistry, Changchun Institute of Applied Chemistry, Renmin St. 5625, Changchun 130022, Jilin, P. R. China
[7]National Synchrotron Light Source-II, Brookhaven National Laboratory, Upton, NY 11973, USA

[\*]Corresponding authors: L.-H. C., liheng.cai@virginia.edu; S. C., chengsh9@msu.edu.

[†]Equal contribution.

# 1. SI Text

## 1.1 Controlled synthesis of reversible polymers

We develop a procedure for synthesizing reversible polymers with precisely controlled molecular architecture (**Fig. S1a,** and **SI Materials and Methods**). We start with quantifying the reaction kinetics of the spacer monomer, hexyl acrylate (HA), and the sticky monomer, 5-acetamido-1-pentyl acrylate (AAPA), using proton nuclear magnetic resonance spectroscopy (see **$^1$H NMR spectra section**). Because both monomers have the same functional group acrylate and similar molecular structure, they exhibit the same reaction rate, as shown by the same fraction of sticky monomers as that of the spacer monomers before and after polymerization (**Fig. S1b**). Moreover, the conversion rate increases linearly with reaction time regardless of the fraction of reversible groups explored in this study (**Fig. S1c**). This informs the controlled synthesis of a series of reversible polymers with fixed molecular weights (MW) while various $\lambda$, the molar fraction of the sticky monomers, ranging 0 to 1 with relatively narrow molecular weight distribution, as shown by gel permeation chromatography (GPC) profiles in **Fig. S1d** and **Fig. S12** and listed in **Table S1**.

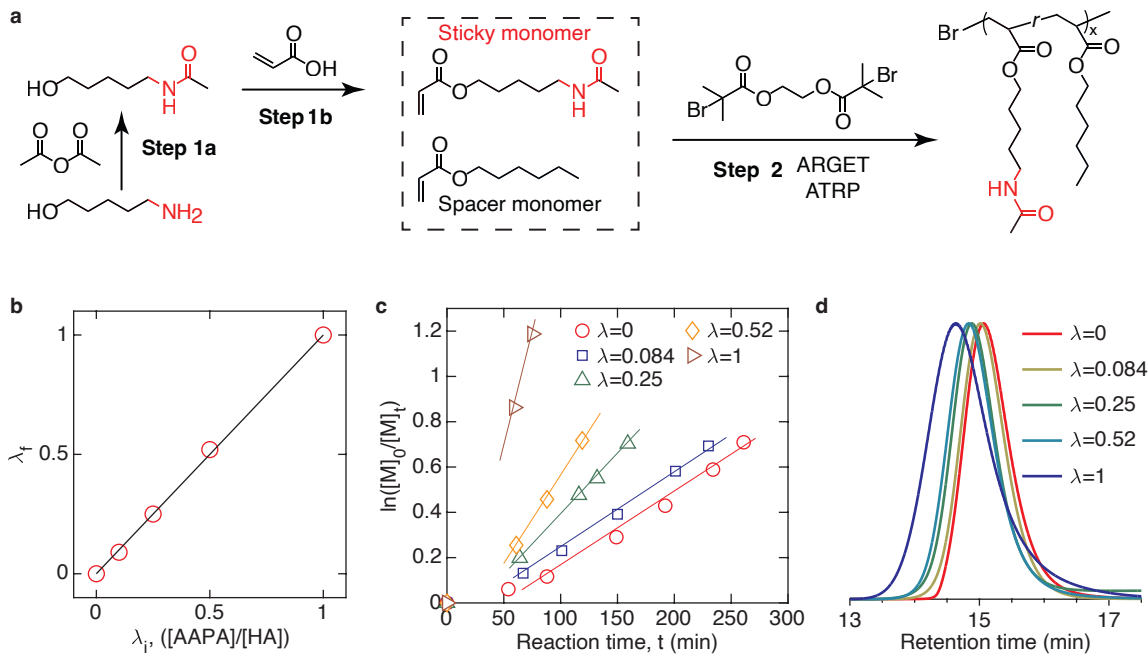

**Figure S1. Synthesis of associative polymers.** (a) Scheme of the synthesis procedure. (b) Dependence of the fraction of AAPA monomers in the final polymer, $\lambda_f$, on the initial feeding ratio, $\lambda_i$, between AAPA and HA monomers. Solid line: $\lambda_f = \lambda_i$. (c) Kinetics of polymerization for the synthesis of reversible polymers with different fractions of stickers. Solid lines are guidance for the eye. (d) GPC traces of unentangled polymers.

## 1.2 Kuhn length of reversible polymers

The spacer and the sticky monomers have the same main-chain C-C bond of length $l = 1.54$Å and the same bond angle $\theta = 68°$. Thus, the Kuhn length $b$ of a reversible polymer is determined through Flory's characteristic ratio $C_\infty$:

$$b = C_\infty l / \cos\left(\frac{\theta}{2}\right) \qquad (S1)$$

Unfortunately, there is no published data on $C_\infty$ for poly(5-acetamido-1-pentyl acrylate) (PAAPA) nor poly(hexyl acrylate) (PHA). Yet, we determine $C_\infty$ based on the molecular characteristics of similar poly(acrylics) and poly(methacrylics) and compare the resulted polymer physical properties such as entanglement modulus against our experiments. For example, poly(hexyl methacrylate) (PHMA), which differs from PHA by having an extra methyl group along the backbone, has $C_\infty^{PHMA} = 13.1$ [1]. Moreover, adding a methyl group increases the $C_\infty^{PMA} = 7.9$ for poly(methyl acrylate) (PMA) to $C_\infty^{PMMA} = 9.0$ for poly(methyl acrylate) (PMMA) [2]. This infers that the $C_\infty$ for PHA likely is lower than that of PHMA by the value of one with $C_\infty = 12.0$.

The addition of an amide group in a AAPA monomer may slightly increase the value of $C_\infty$ for polymers with high fractions of stickers. But since the amide group is at the end of the side group, there is more space available compared to the location closer to the polymer backbone, such that the geometric hindrance is alleviated. Therefore, we use $C_\infty = 12.0$ for all reversible polymers. This gives the length of a Kuhn segment $b = 22$ Å [see eq. (S1)], and the number of main-chain C-C bonds per Kuhn segment is $N_k = b/\left[l \cos\left(\frac{\theta}{2}\right)\right] = C_\infty/\cos^2\left(\frac{\theta}{2}\right) = 17.4$. The mass of a Kuhn segment for a reversible polymer with sticky monomers of a fraction $\lambda$ is:

$$M_0(\lambda) = \frac{1}{2} N_k [m_{HA}(1-\lambda) + m_{AAPA}\lambda] \tag{S2}$$

in which $m_{HA} = 156$ Da is the mass of the spacer monomer HA, $m_{AAPA} = 199$ Da is the mass of the sticky monomer AAPA, and the factor 1/2 comes from that each acrylic monomer has two main-chain C-C bonds.

We calculate the entanglement MW of a flexible linear polymer, $M_e$, based on the Kavassalis−Noolandi conjecture [3]. The criterion is that the overlap parameter for an entanglement strand is constant regardless of polymer species, $P_e \equiv \frac{a^3}{v_0 N_e} = 20$, where $a = bN_e^{1/2}$ is the size of an entanglement strand, $N_e$ is the number of Kuhn segments per entanglement strand, and $v_0 = M_0/\rho N_{Av}$ is the volume of a Kuhn segment with $\rho$ being the polymer density and $N_{Av}$ being the Avogadro's number. Defining the packing length as the ratio of the occupied volume of a Kuhn segment to the mean-square end-to-end distance, $p \equiv v_0/b^2$, one obtains:

$$M_e = M_0 N_e = M_0 \left(\frac{pP_e}{b}\right)^2 \tag{S3}$$

And the entanglement modulus is:

$$G_e = k_B T \rho N_{Av}/M_e \tag{S4}$$

where $k_B$ is Boltzmann constant, $T$ is the absolute temperature, $\rho = 1.04$ g/cm³ is the density of PHA. For PHA, $M_0 = 1360$ Da, $v_0 = 2170$ Å³, $p = 4.5$ Å, $M_e = 23$ kDa, $N_e = 17$, and $M_e = 110$ kPa.

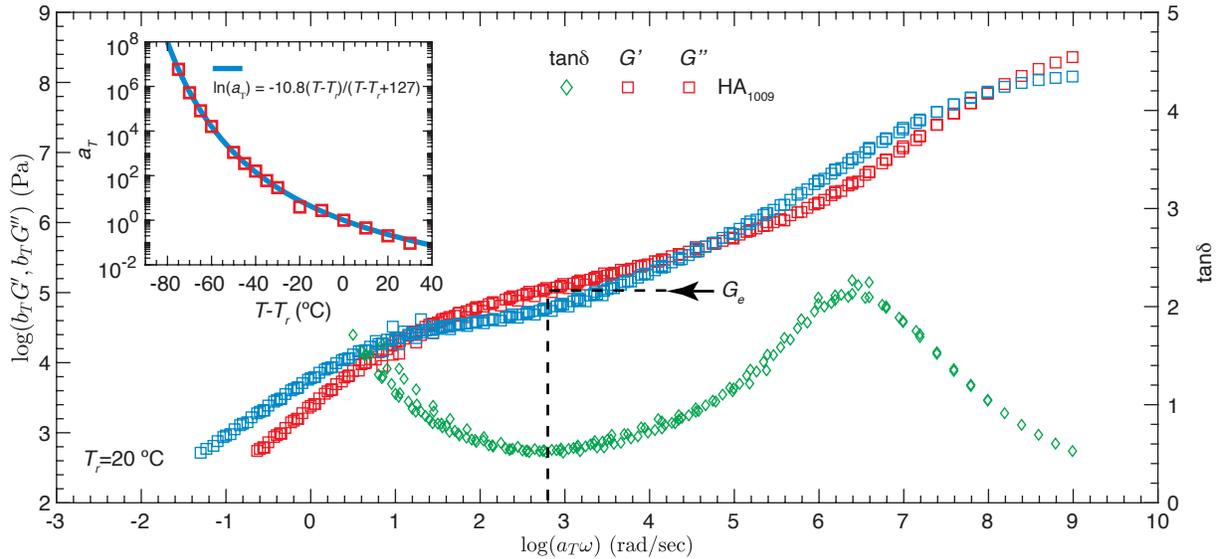

**Figure S2. Linear viscoelasticity of an entangled poly(hexyl acrylate) melt.** Master curves of storage ($G'$) and loss ($G''$) moduli, as well as loss factor $\tan\delta \equiv G''/G'$ as a function of angular frequency $\omega$ for the melt of entangled PHA at a reference temperature $T_r = 20$°C. The modulus scale shift factor $b_T = \rho T/\rho_r T_r$, where $\rho_r$ is the polymer density at the reference temperature $T_r$, and $T$ is the absolute temperature at which the oscillatory frequency measurements are performed. Inset: The time scale shift factor $a_T$ can be described by Williams–Landel–Ferry (WLF) equation: $\ln a_T = -10.8 \frac{T-T_r}{T-T_r+127}$ (solid line).

We compare the calculated entanglement modulus to the experimentally measured value. To do so, we synthesize a long PHA polymer with degree of polymerization (DP) of 1009, as confirmed by proton NMR and GPC measurements (**Fig. S10** and **SI $^1$H NMR spectra**). The corresponding molecular weight, 157 kDa, is well above the critical MW 46 kDa to ensure the formation of an entangled polymer melt. Following a similar approach (see section on **Rheometry** in **SI Materials and Methods**), we perform an oscillatory shear frequency sweep of the entangled polymer melt at various temperatures, based on which we construct a master curve for the linear viscoelasticity of the polymer, as shown by the squares in **Fig. S2** with the time scale shift factor $a_T$ plotted in the inset. This allows us to determine the entanglement modulus of $G_e \approx 110$ kPa, which corresponds to the local minimum of loss $\tan \delta = G'/G''$, as indicated by the dashed lines in **Fig. S2**. The experimentally measured $G_e$ is nearly identical to the theoretical estimate, validating the choice of Flory's characteristic ratio $C_\infty = 12.0$ for PHA and reversible polymers.

**1.3 Reversible interactions increase the polymer glass transition temperature**

The dynamic mechanical properties of polymers depend on the measurement temperature relative to their glass transition temperature, $T_g$. Therefore, it is critical to determine the $T_g$ of reversible polymers. We use differential scanning calorimetry (DSC) to quantify the $T_g$ for both unentangled and entangled polymers (**Fig. S3a**). The $T_g$ is defined as the deflection point for the dependence of heat flow on temperature ramp, as shown by the peaks in **Fig. S3b**. The measured $T_g$ for pure PHA$_{254}$ is 210 K, close to the literature value 216 K [4]. Such a small difference is likely because the MW of PHA$_{254}$ is not large enough for $T_g$ to reach the maximum, $T_{g,\infty}$, as suggested by the classical Flory-Fox equation [5], $T_g(M_n) = T_{g,\infty} - C/M_n$, where $C$ is constant with typical values of $10^4$-$10^5$ K g/mol (Chapter 12, in ref. [6]) and $M_n$ is the number-average molecular weight of polymers. For the reversible polymers explored in this study, the MW is greater than 30 kDa, such that the value of $C/M_n$ is on the order of 1 K and thus negligible compared to the polymer glass transition temperature. For example, for a fixed $\lambda$, $T_g$ is is nearly the same for polymers with different MWs, as shown by the overlap between circles for unentangled and squares for entangled polymers in **Fig. 2**.

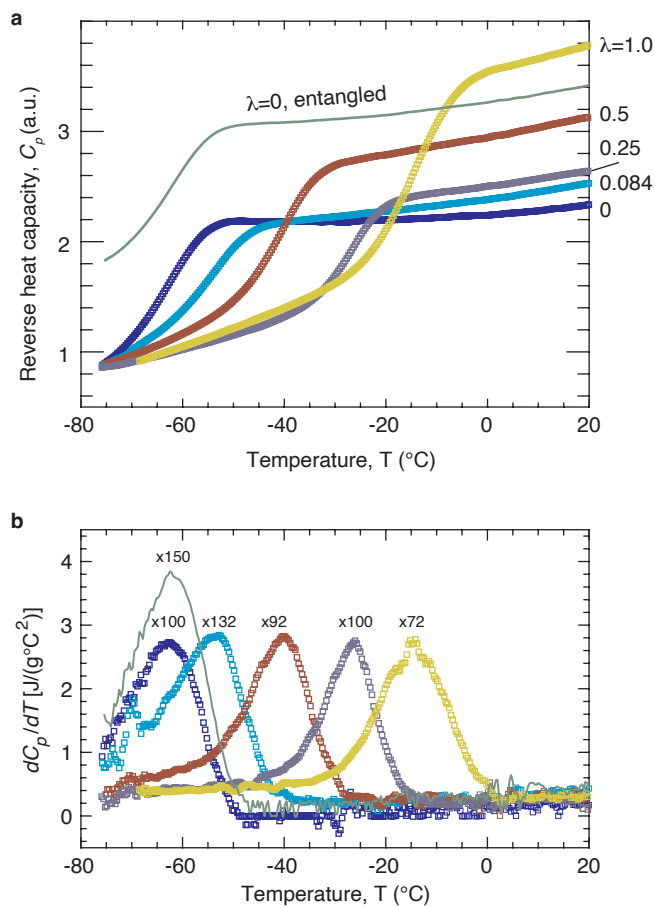

**Figure S3. The glass transition temperature of reversible polymers.** (a) Depdence of reverse heat capacity $C_p$ of unetangled (squares) and entangled (solid line) reversible polymers. The curves are shifted vetically for clarity. a.u.: arbitrary units. (b) Derivative of the heat capacity versus temperature. The value of $dC_p/dT$ for each sample is multiplied by a factor for clarity, and the values of the factor are listed near the peak of each curve. The peak corresponds to the glass transition temperature of a polymer. Symbols from left to right are for unentangled polymers with $\lambda = 0, 0.084, 0.25, 0.52, 1.0$; solid line is for entangled control polymer with $\lambda = 0$.

The glass transition temperature of reversible polymers increases significantly with the fraction of stickers. For example, as $\lambda$ increases from 0 to 1, $T_g$ increases by nearly 50 degrees from $-64.5$ °C to $-16.5\,°\text{C}$; this is consistent with the understanding that stronger inter- and intra-molecular interactions result in relatively higher $T_g$. Importantly, at the highest fraction of reversible bonds with $\lambda = 1$, the $T_g = -16.5\,°\text{C}$ remains well below room temperature, highlighting the potential of using these reversible polymers to create supramolecular elastomers. Collectively, our results show that the $T_g$ of reversible polymers is determined by not much the MW but the fraction of reversible bonds.

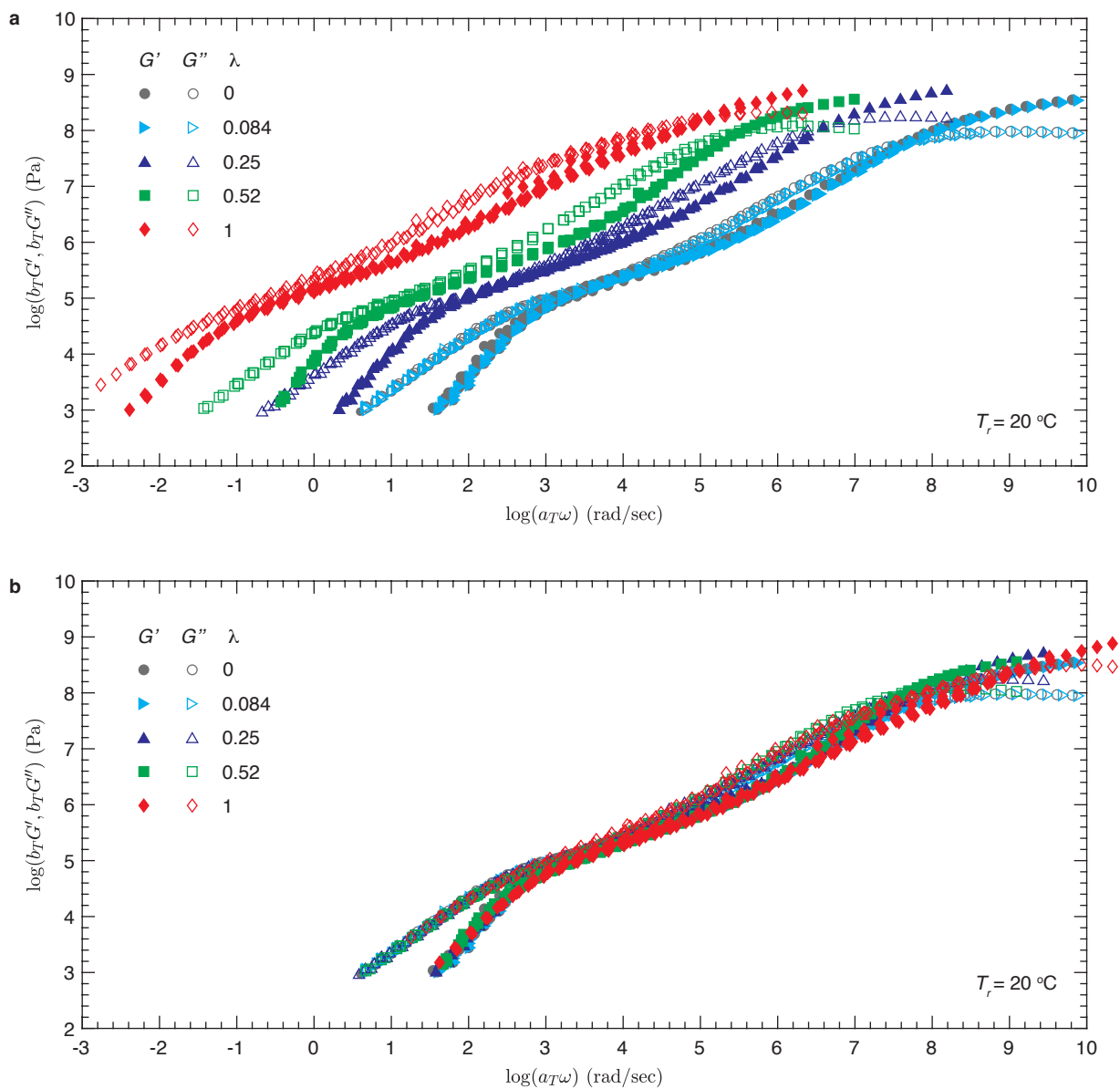

**Figure S4. Linear viscoelasticity of associative polymers.** (a) Master curves of associative polymers with various concentrations of stickers at reference temperature $T_r = 20°C$. These are the same as those in Fig. 3 but placed here as a reference for (b), which shows that the master curves for associative polymers with $\lambda \neq 0$ can be shifted horizontally to overlap with the viscoelastic master curve of $\lambda = 0$.

### 1.4 Shapes of linear viscoelastic spectra for associative polymers

To explore the effects of reversible interactions on the shape of the linear viscoelastic spectra, we horizontally shift the master curves for associative polymers with $\lambda \neq 0$ to overlay with the terminal relaxation moduli for the control polymer with $\lambda = 0$. Interestingly, the master curves for all associative polymers nearly perfectly overlay with each other at the intermediate frequency and the low frequency regions regardless of the concentration of stickers, as shown in **Fig. S4b**. This behavior suggests that reversible interactions significantly slow down the polymer dynamics but nearly do not change the shape of the linear viscoelastic spectra of associative polymers.

### 1.5 Glassy relaxation of associative polymers

At time scales shorter than the segmental relaxation time, the polymer relaxation transitions from Rouse to glassy dynamics. Interestingly, as $\lambda$ increases from 0 to 1, the width of the transition zone increases by nearly one decade. This increase is further confirmed by applying Kohlrausch–Williams–Watts (KWW) model [7–9], a phenomenological stretched exponential form, to fit the polymer glass dynamics [see **Table S1** for fitting parameters]:

$$G_g(t) = G_g(0) \exp\left[-\left(\frac{t}{\tau_{KWW}}\right)^\beta\right] \tag{S5}$$

in which $G_g(0)$ is the glassy modulus at $t \to 0$, $\tau_{KWW}$ is the characteristic time of the glassy relaxation, and the exponent $\beta$ describes the distribution of glassy relaxation modes. The lower the value of $\beta$, the broader the mode distribution. As $\lambda$ increases from 0 to 0.5, the value of $\beta$ decreases from 0.23 to 0.20; further increasing $\lambda$ to 1 leads to a very broad transition with a small value $\beta = 0.15$ (see **Table S1**). Collectively, our results show that the addition of reversible bonds broadens the glass transition zone.

**1.6 Atomistic simulation of melts of Kuhn monomers**

To study the dynamics on the length scale of Kuhn monomers, we perform atomistic simulations of 9-mers, which are the molecules each consisting of 9 chemical monomers and comparable to the Kuhn monomers in the experiments. The classical OPLS All-Atom force field [10] is employed. The interaction parameters for the inter-atomic Lennard-Jones (LJ) potentials, covalent bonding, bond bending, and torsional potentials are selected to match the local chemical environments of different atoms in the associative polymers. The cut-off distance for the LJ potentials was set as 12 Å. The point charges carried by the atoms in a molecule are arranged to yield a zero net charge. The long-range electrostatic interaction was computed using a particle-particle particle-mesh (PPPM) solver with numerical precision $10^{-4}$. Depending on whether the amide group is present, a chemical monomer is either sticky or non-sticky. Following the experiments, the number of sticky monomers per 9-mer $n$ = 0, 1, 2, 3, 6, and 9 in the simulations, corresponding to the fraction of sticky monomers $\lambda$ =0, 0.11, 0.22, 0.33, 0.67, and 1. Snapshots of 9-mers with $\lambda$ = 0 and 1 in the simulations are shown in **Fig. S5a, b**.

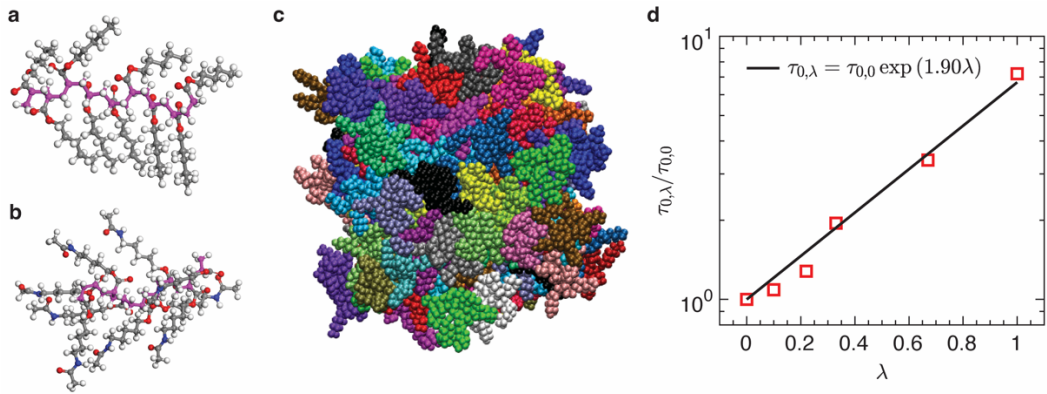

**Figure S5. Atomistic simulation of melts of Kuhn monomers.** Simulation snapshots of (a) a 9-mer with $\lambda = 0$, (b) a 9-mer with $\lambda = 1$, where magenta, grey, white, red, and blue spheres indicate C in the backbone, C in the side groups, H, O, and N, respectively, and (c) a melt sample of 125 9-mer molecules. (d) Diffusion time $\tau_{0,\lambda}$ of a 9-mer molecule as a function of the fraction of sticky monomers $\lambda$ in the atomistic simulations at T=500K. Simulation data points (squares) are fit to $\tau_{0,\lambda} = \tau_0 \exp(c\lambda)$. The fitting result is $c = 1.90 \pm 0.08$ with the best fit shown by the solid line.

For each λ, we built a sample of 125 molecules in a cubic simulation box with periodic boundary conditions, as illustrated in **Fig. S5c**. All the samples were equilibrated at temperature T = 500 K and pressure P = 1 atm using the Nosé–Hoover thermostat and Nosé–Hoover barostat with characteristic times of 50 fs and 500 fs, respectively. The NPT simulation lasted for 1 ns with a timestep 1 fs. The mass density ρ and the simulation box size L for different samples at the thermal equilibrium are listed in **Table S3**. After the equilibration run, the diffusion of 9-mers in the melt state was simulated at constant volume and temperature T=500 K using the Nosé–Hoover thermostat with a characteristic time of 50 fs. The NVT diffusion simulation lasted for 10 ns with a timestep 1 fs. The center-of-mass mean squared displacement (MSD) of 9-mers was obtained as a function of time t. The diffusion coefficient of a 9-mer molecule was calculated as $D = \lim_{t\to\infty} MSD/6t$. We determine the diffusion time of a 9-mer molecule as $\tau_{0,\lambda} = l^2/6D$, where $l$ is the average size of a 9-mer. The value of l is estimated as $l = L/\sqrt[3]{125} = L/5$, which increases from 14.2 Å at $\lambda = 0$ to 14.5 Å at $\lambda = 1$ by only 2%. The plot of $\tau_{0,\lambda}/\tau_{0,0}$ as a function of $\lambda$ is shown in **Fig. S5d**. All the atomistic samples were prepared with the aid of Moltemplate [11] and simulated using the simulation package LAMMPS [12].

**Table S1. Molecular parameters of associative polymers.**
$x$, number of HA monomers; $y$, number of AAPA monomers; PDI, polydispersity index; $\lambda$, fraction of AAPA monomers; $M_0$, mass of a Kuhn segment; $M_n$, number average mass of a polymer determined by $^1$H NMR; $\tau_{s0}$, relaxation time of a *renormalized* Kuhn segment at room temperature; $G_g(0)$, glassy modulus at $t \to 0$; $\beta$, stretching exponent for KWW model; $\tau_{KWW}$, characteristic time of the glassy relaxation in KWW model.

| Sample | HA$_x$-r-AAPA$_y$ | | | | | | rRouse | KWW | | |
|---|---|---|---|---|---|---|---|---|---|---|
| | $x$ | $y$ | PDI | $\lambda$ | $M_0$ (Da) | $M_n$ (Da) | $\tau_{s0}$ (10$^{-7}$ sec) | $G_g(0)$ (GPa) | $\beta$ | $\tau_{KWW}$ (10$^{-11}$ sec) |
| 1 | 254 | 0 | 1.15 | 0 | 1359 | 39680 | 3.2 | 0.84 | 0.23 | 3.3 |
| 2 | 229 | 21 | 1.14 | 0.084 | 1391 | 39959 | 3.7 | 0.9 | 0.22 | 1.9 |
| 3 | 189 | 64 | 1.18 | 0.25 | 1454 | 42278 | 5.7×10$^1$ | 2.3 | 0.21 | 1.1×10$^1$ |
| 4 | 123 | 133 | 1.18 | 0.52 | 1554 | 45715 | 4.9×10$^2$ | 2.6 | 0.2 | 6.5×10$^1$ |
| 5 | 0 | 280 | 1.61 | 1 | 1733 | 55790 | 2.1×10$^4$ | 3.5 | 0.15 | 8.9×10$^1$ |
| 6 | 1009 | 0 | 1.5 | 0 | 1359 | 157626 | | | | |

**Table S2. Summary of synthesis conditions of all polymers.**
Catalyst is made by dissolving Me$_6$TREN (92 mg, 0.4 mmol) and CuCl$_2$ (5.4 mg, 0.04 mmol) in 1 mL DMF.

| Sample | HA (mmol) | AAPA (mmol) | Initiator (mmol) | Catalyst (μL) | Sn(EH)$_2$ (mmol) | Anisole (mL) | DMF (mL) | Temperature (°C) | Time (min) | Conv. (%) | $x$ | $y$ | $\lambda$ |
|---|---|---|---|---|---|---|---|---|---|---|---|---|---|
| 1 | 76.8 | 0 | 0.154 | 384 | 0.307 | 16 | 0 | 70 | 261 | 50.8 | 254 | 0 | 0 |
| 2 | 28 | 3.11 | 0.062 | 156 | 0.125 | 6.2 | 0 | 70 | 230 | 52.4 | 229 | 21 | 0.084 |
| 3 | 22.4 | 7.53 | 0.060 | 150 | 0.120 | 6 | 0 | 70 | 159 | 50.5 | 189 | 64 | 0.25 |
| 4 | 11.3 | 11.3 | 0.045 | 113 | 0.091 | 4.6 | 0 | 70 | 119 | 51.2 | 123 | 133 | 0.52 |
| 5 | 0 | 14.1 | 0.035 | 73 | 0.058 | 0 | 4.4 | 70 | 75 | 69.5 | 0 | 280 | 1 |
| 6 | 32 | 0 | 0.016 | 320 | 0.128 | 6.4 | 0 | 80 | 304 | 50.5 | 1009 | 0 | 0 |

**Table S3. Parameters for the equilibrated melt samples in the atomistic simulations.**
Each sample contains 125 molecules with 9 chemical monomers per molecule.

| $\lambda$ | 0 | 0.11 | 0.22 | 0.33 | 0.67 | 1 |
|---|---|---|---|---|---|---|
| $\rho$ (g/cm$^3$) | 0.82 | 0.84 | 0.85 | 0.87 | 0.93 | 0.98 |
| $L$ (Å) | 71.1 | 71.1 | 71.2 | 71.4 | 72.0 | 72.3 |

## 2. SI Materials and Methods

### 2.1 Materials

Hexyl acrylate (98%), acetic anhydride (⩾98%), 5-amino-1-pentanol (⩾92%), acrylic acid (99%), $K_2CO_3$ (⩾99%), *N*-(3-Dimethylaminopropyl)-*N*'-ethylcarbodiimide hydrochloride (EDC, 98%), *N,N*-diisopropylethylamine (>99%), Copper(II) chloride ($CuCl_2$, 99.999%), *tris*[2-(dimethylamino)ethyl]amine ($Me_6TREN$), ethylene bis(2-bromoisobutyrate) (2f-BiB, 97%), Tin(II) 2-ethylhexanoate ($Sn(EH)_2$, 92.5–100%), anisole (⩾99.7%) were purchased from Sigma Aldrich and used as received. Methanol (Certified ACS), ethyl acetate (Certified ACS), hexanes (Certified ACS), dichloromethane (Certified ACS), diethyl ether (Certified ACS), dimethylformamide (DMF, Certified ACS), tetrahydrofuran (THF, Certified ACS) and THF (HPLC), were purchased from Fisher and used as received.

### 2.2 Polymer synthesis and characterization

For the synthesis of reversible polymers, we use activator regenerated by electron transfer (ARGET) atom transfer radical polymerization (ATRP) [13]. We synthesize the polymers by copolymerizing hexyl acrylate (HA) with 5-acetamidopentyl acrylate (AAPA), which carries an amide group at one of its two ends and serves as the sticky monomer. The reaction conditions for the synthesis of all polymers are summarized in **Table S2**. Below we describe the detailed synthesis protocols.

*Step I. Synthesis of 5-acetamido pentyl acrylate (AAPA).* First, a flask is charged with 5-amino-1-pentanol (25 g, 242.3 mmol) and ethyl acetate (250 mL). Acetic anhydride (28.1 g, 275.4 mmol) is added dropwise with vigorous stirring under nitrogen. After finishing the addition of acetic anhydride, the reaction mixture is stirred at room temperature for 2 hours followed by the addition of methanol (80 mL) and $K_2CO_3$ (28 g, 202.6 mmol). The mixture is vigorously stirred for another 15 min followed by the filtration of undissolved solid if any. The filtered solution is concentrated by a rotary evaporator (Buchi R-205) to obtain 5-acetamido-1- pentanol (30.5 g) with a yield of 87.1%. The success for the synthesis is confirmed by $^1$H NMR (600 MHz, CDCl$_3$) $\delta$=3.53 (t, 2H), 3.14 (q, 2H), 1.89 (s, 3H), 1.47 (m, 4H), 1.32 (m, 4H).

Second, a flask is charged with 5-acetamido-1- pentanol (3.34 g, 23.0 mmol), acrylic acid (2.48 g, 34.5 mmol), EDC (7.27 g, 37.9 mmol), *N,N*-diisopropylethylamine (4.9 g, 37.9 mmol) and dichloromethane (100 mL). The reaction is stirred at room temperature for 48 h under nitrogen. We dilute the reaction mixture by adding another 100 mL dichloromethane, and sequentially wash the solution using an aqueous solution of NaOH (1.0 M, 100 mL), an aqueous solution of HCl (1.0 M, 100 mL), a saturated aqueous solution of NaHCO$_3$ (150 mL), and a saturated aqueous solution of NaCl (150 mL). This washing process yields an organic supernatant, which is dried with Na$_2$SO$_4$ for 12 h and then concentrated by a rotary evaporator to obtain the crude product. We purify the crude product by passing it through a silica column using ethyl acetate/hexanes = 1/9 (v/v) as eluent. 5-acetamido pentyl acrylate (AAPA) (3.6 g) is obtained with a yield of 78.6%. $^1$H NMR (600 MHz, CDCl$_3$) $\delta$=6.37 (d, 1H), 6.11 (dd, 1H), 5.82 (d, 1H), 5.58 (s, 1H), 4.15 (t, 2H), 3.23 (q, 2H), 1.96 (s, 3H), 1.68 (m, 2H), 1.53 (m, 2H), 1.39 (m, 2H).

*Step II-a. Synthesis of the control polymer without reversible groups.* A 25 mL Schlenk flask is charged with 2f-BiB (55.3 mg, 0.154 mmol), hexyl acrylate (12 g, 76.8 mmol) and anisole (16 mL). We dissolve Me$_6$TREN (92 mg, 0.4 mmol) and CuCl$_2$ (5.4 mg, 0.04 mmol) in 1 mL DMF to make a catalyst solution. Then, we add 384 μL catalyst solution, containing $1.54\times10^{-1}$ mmol Me$_6$TREN and $1.54\times10^{-2}$ mmol CuCl$_2$, to the mixture and bubble it with nitrogen for 35 min to remove oxygen. Afterward, we use a glass pipet to quickly add the reducing agent, Sn(EH)$_2$ (124.5 mg, 0.307 mmol) in 300 μL anisole, to the reaction mixture. We seal the flask and then immerse it in an oil bath at 70°C to start the reaction. We monitor the reaction by taking out a small amount of mixture at different time points to determine the conversion using proton NMR and stop the reaction after 261 min. Based on $^1$H NMR, the conversion is 50.8% and DP is 254.

The rest reaction mixture is diluted with THF and passed through a neutral aluminum oxide column to remove the catalyst. The collected solution is concentrated by a rotary evaporator. We use methanol to precipitate the polymer, re-dissolve the sediment in THF to make a homogenous solution, and repeat this precipitation procedure another 2 times to ensure that all unreacted monomers and impurities are completely removed. After purification, the sample is dissolved in THF and transferred to a glass vial and dried in the hood for 16 h, and then transferred to a vacuum oven (Thermo Fisher, Model 6258) at room temperature for 24 h to completely remove the solvent.

*Step II-b. Example synthesis of reversible polymer consisting of both HA and AAPA monomers.* A 25 mL Schlenk flask is charged with 2f-BiB (22.4 mg, 0.062 mmol), hexyl acrylate (4.38 g, 28.0 mmol), AAPA (0.62 g, 3.11 mmol) and anisole (6.2 mL). We dissolve Me$_6$TREN (92 mg, 0.4 mmol) and CuCl$_2$ (5.4 mg, 0.04 mmol) in 1 mL DMF to make a catalyst solution. Then, we add 156 μL of catalyst solution, containing 6.2×10$^{-2}$ mmol Me$_6$TREN and 6.2×10$^{-3}$ mmol CuCl$_2$, to the mixture and bubble it with nitrogen for 30 min to remove oxygen. Afterward, the reducing agent, Sn(EH)$_2$ (50.4 mg, 0.125 mmol) in 200 μL anisole, is quickly added to the reaction mixture using a glass pipet. We seal the flask and then immerse it in an oil bath at 70°C to start the reaction. The reaction is monitored by taking out a small amount of mixture at different time points to determine the conversion using proton NMR and stopped after 230 min. From proton NMR, the conversion is 50.0% and the total DP is 250. The purification procedure is the same as the synthesis of the control middle block. After purification, from $^1$H NMR, the DP of hexyl acrylate is 229, the DP of AAPA is 21, the percentage of reversible groups is 8.4%.

*Step II-c. Synthesis of a reversible polymer with AAPA sticky monomer only.* A 25 mL Schlenk flask is charged with 2f-BiB (12.6 mg, 0.035 mmol), AAPA (2.9 g, 14.6 mmol), and DMF (4.4 mL). We dissolve Me$_6$TREN (92 mg, 0.4 mmol) and CuCl$_2$ (5.4 mg, 0.04 mmol) in 1 mL DMF to make a catalyst solution. Then, we add 73 μL catalyst solution, containing 2.9×10$^{-2}$ mmol Me$_6$TREN and 2.9×10$^{-3}$ mmol CuCl$_2$, to the mixture and bubble it with nitrogen for 30 min to remove oxygen. Afterward, the reducing agent, Sn(EH)$_2$ (23.6 mg, 0.058 mmol) in 200 μL anisole, is quickly added to the reaction mixture using a glass pipet. We seal the flask and then immerse it in an oil bath at 65°C to start the reaction. The reaction is monitored by taking out a small amount of mixture at different time points to determine the conversion using $^1$H NMR and stopped after 75 min. From $^1$H NMR, the conversion is 69.5% and the DP is 280. The purification procedure is similar to the synthesis of the control middle block. The only difference is that diethyl ether is used for precipitation, which can dissolve AAPA but is a poor solvent for PAAPA. The purification is repeated 3 times to ensure that all unreacted monomers and impurities are removed. After purification, the sample is dissolved in methanol and transferred to a glass vial and dried in the hood for 16 h, and then transferred to a vacuum oven (Thermo Fisher, Model 6258) at room temperature for 48 h to completely remove the solvent.

## 2.3 Differential Scanning Calorimetry (DSC)

We determine the glass transition temperature $T_g$ of all samples using a temperature modulated differential scanning calorimeter (TMDSC) DSC250 (TA instruments) from 308K to 193K at a cooling rate of 2 K/min with a modulation rate of 1 K/min and a modulation frequency of 60 Hz. Before characterization all the samples are dried at room temperature (~293 K) under vacuum (30 mbar) for at least 24 hours. A standard aluminum DSC pan is used for all the measurements. The absolute specific heat capacity, $C_p$, is determined through measurements first the empty pan (the calibration line) and then the same pan with 10 mg samples following the same cooling protocol. The $T_g$ values are then determined as the peak of the derivative of the heat capacity versus temperature (see **Fig. S3b**).

## 2.4 Rheometry

We use a stress-controlled rheometer (Anton Paar MCR302) equipped with an environmental chamber (CTD600) or a Peltier (PTD200) to quantify the linear viscoelasticity of associative polymers. Both the environmental chamber and the Peltier have a temperature accuracy of $\pm 0.01\ K$. The chamber is filled with N$_2$, an inert gas to protect the polymer from moisture that may influence the hydrogen bond strength. For each sample, we use parallel plates perform oscillatory shear frequency measurements over a wide range of temperatures from the corresponding $T_g$ to $T_g$+80K. Two types of parallel plates are used in the measurements. For temperatures close to $T_g$ with modulus above $10^7$ Pa, we use parallel plate with 4 mm in diameter and vary the strain amplitude from 0.01% to 0.1%. For temperature above $T_g$ with modulus below $10^7$ Pa, we use a parallel plate of 8 mm in diameter and increase the strain amplitude from 0.1% to 3%. In all configurations, we keep the sample thickness at 1 mm and set the angular frequency in the range of $10^2$ -$10^{-1}$ rad/s. For the polymer with the highest fraction of amide groups, which is expected to be the most sensitive to moisture and measurement conditions, independent measurements show nearly identical results (**Fig. S11**).

## 2.5 Small angle X-ray scattering

We use the Soft Matter Interfaces (12-ID) beamline [14] at the Brookhaven National Laboratory to perform SAXS measurements on annealed bulk polymers. The sample-to-detector distance is 8.3m, and the radiation wavelength is $\lambda = 0.77$ Å. The scattered X-rays are recorded using an in-vacuum Pilatus 1M detector, consisting of 0.172 mm square pixels in a 941×1043 array. The raw SAXS images are converted into *q*-space, visualized in Xi-CAM software [15], and radially integrated using a custom Python code. The one-dimensional intensity profile, *I(q)*, is plotted as a function of the scattering wave vector, $|\vec{q}| = q = 4\pi\lambda^{-1}\sin(\theta/2)$, where $\theta$ is the scattering angle.

## 2.6 $^1$H NMR characterization

We use $^1$H NMR to determine the conversion of HA and AAPA. Chemical shifts for $^1$H NMR spectra are reported in parts per million compared to a singlet at 7.26 ppm in CDCl$_3$ or a singlet at 2.50 ppm in DMSO-$d_6$.

*Conversion of HA for the control polymer.* The conversion is calculated based on the conversion of HA monomers to the polymer poly(hexyl acrylate) (PHA), which is measured by the NMR spectra of the raw reaction mixture, as shown in **Fig. S6**. The area of peak **a** at 4.19 ppm, $A_{HA}$, corresponds to two H on the methylene group connected with the oxygen atom in hexyl acrylate monomer. The area of peak **a'** at 4.07 ppm, $A_{PHA}$, corresponds to two H on the methylene group connected with the oxygen atom in hexyl acrylate repeating unit of PHA. The conversion of HA equals $A_{PHA} \times 100\% / (A_{HA} + A_{PHA})$. In **Fig. S6**, the conversion of HA is $1.034 \times 100\% / (1 + 1.034)$ = 50.8%. Because the molar ratio between HA monomer and initiator is 500, the DP is 500 × 50.8% = 254.

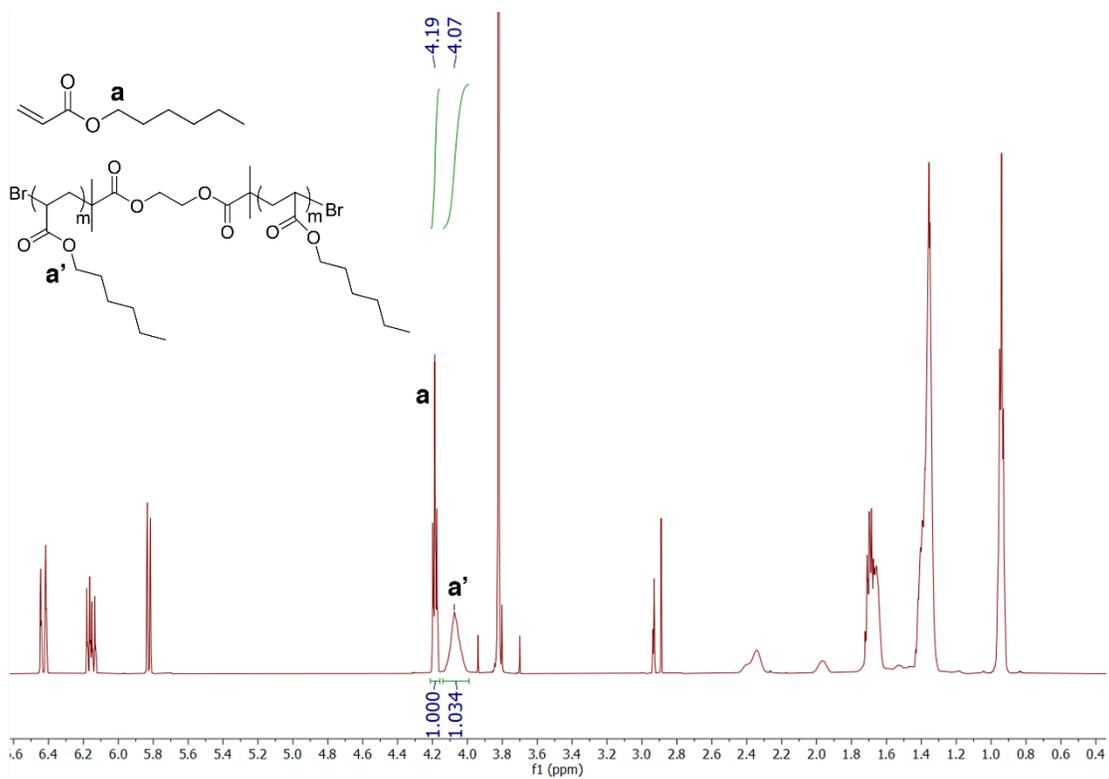

**Figure S6.** $^1$H NMR spectra of raw mix of ARGET ATRP of hexyl acrylate in CDCl$_3$.

*Example of the conversion of HA and AAPA for a reversible polymer.* The total conversion is calculated based on the conversion of HA and AAPA to the random copolymer poly(HA-*r*-AAPA), which is measured by the NMR spectra of the raw reaction mixture, as shown in **Fig. S7**. The area of peak **a** $A_{HA}$ and peak **b** $A_{AAPA}$ at 4.18 ppm corresponds to two H on the methylene group connected with the oxygen atom in HA and AAPA monomers, respectively. The area of peak **a'** $A_{PHA}$ and **b'** $A_{PAAPA}$ at 4.06 ppm corresponds to two H on the methylene group connected with the oxygen atom in HA and AAPA repeating units, respectively. The total conversion of HA and AAPA equals to $(A_{PHA}+A_{PAAPA}) \times 100\% / (A_{HA}+A_{PHA}+A_{AAPA}+A_{PAAPA})$. The total conversion of HA and AAPA in **Fig. S7** equals to $1 \times 100\% / (1+1) = 50.0\%$. For this polymerization, the molar ratio between HA and AAPA and initiator is 500. Therefore, the total DP is $500 \times 50.0\% = 250$.

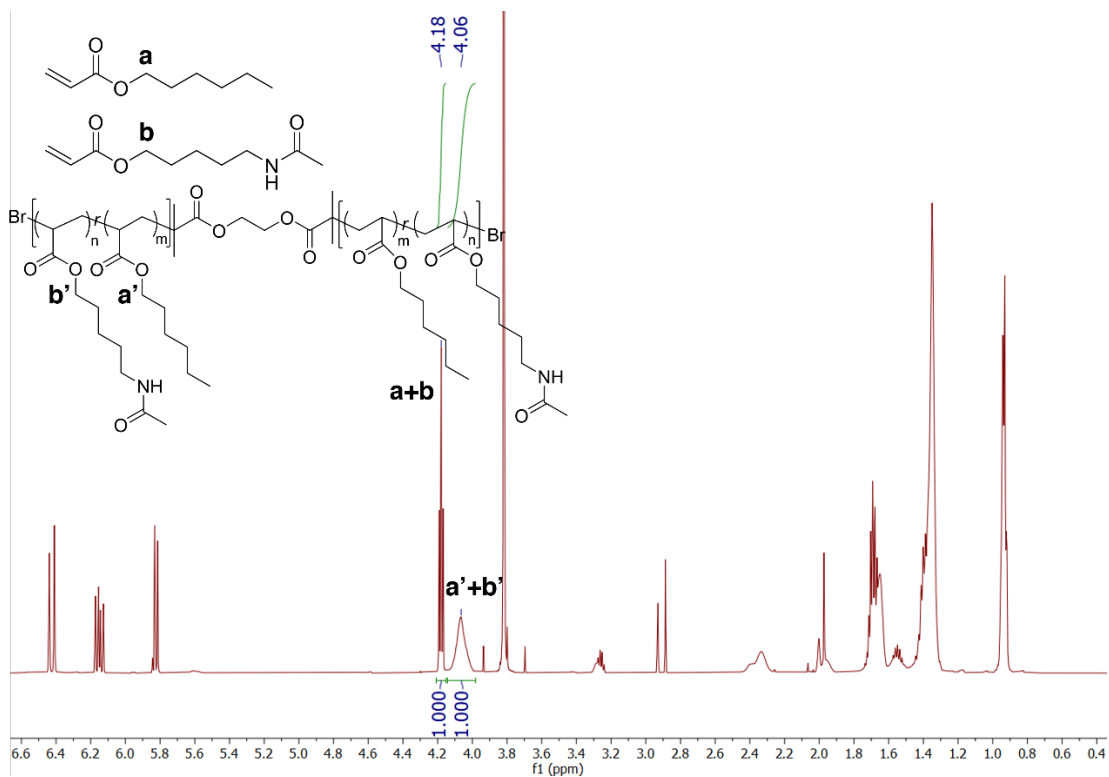

**Figure S7.** $^1$H NMR spectra of raw mix of ARGET ATRP of HA and AAPA in CDCl$_3$.

*Example of the conversion of AAPA for reversible polymer with sticky monomer only.* The total conversion is calculated based on the conversion of AAPA to the polymer PAAPA, which is measured by the NMR spectra of the raw reaction mixture, as shown in **Fig. S8**. The area of peak **a** $A_{AAPA}$ at 4.09 ppm corresponds to two H on the methylene group connected with the oxygen atom in AAPA monomers. The area of peak **a'** $A_{PAAPA}$ at 3.95 ppm corresponds to two H on the methylene group connected with the oxygen atom in AAPA repeating units. The total conversion of AAPA equals to $A_{PAAPA} \times 100\% / (A_{AAPA}+A_{PAAPA})$. The total conversion of AAPA in **Fig. S8** equals to $2.279 \times 100\% / (1+2.279) = 69.5\%$. For this polymerization, the molar ratio between AAPA and initiator is 403. Therefore, the total DP of AAPA is $403 \times 69.5\% = 280$.

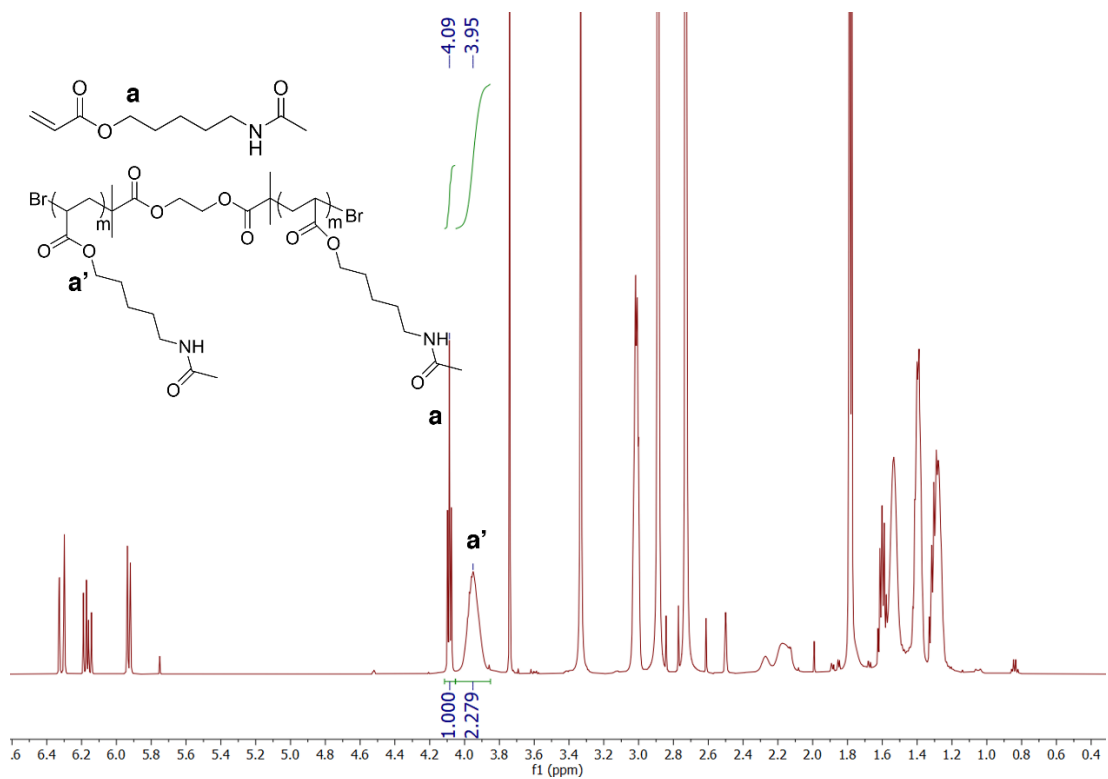

**Figure S8.** $^1$H NMR spectra of raw mix of ARGET ATRP of AAPA in DMSO-$d_6$.

*Example for the calculation of the fraction of reversible groups in a reversible polymer.* In **Fig S9**, the area of peak **a** $A_{PHA-O}$ and peak **b** $A_{PAAPA-O}$ at 3.99 ppm corresponds to two H on the methylene group connected with the oxygen atom in HA and AAPA repeating units, respectively. The area of peak **c** $A_{PAAPA-N}$ at 3.22 ppm corresponds to two H on the methylene group connected with the nitrogen atom in AAPA repeating units. The fraction of reversible groups equals to $A_{PAAPA-N} \times 100\% / (A_{PHA-O}+A_{PAAPA-O})$. The fraction of reversible groups in **Fig. S9** equals to $0.084 \times 100\% /1 = 8.4\%$. For this copolymer, the total DP for HA and AAPA is 250. Therefore, the DP of AAPA is $250 \times 8.4\% = 21$, the DP of HA is $250-21 = 229$.

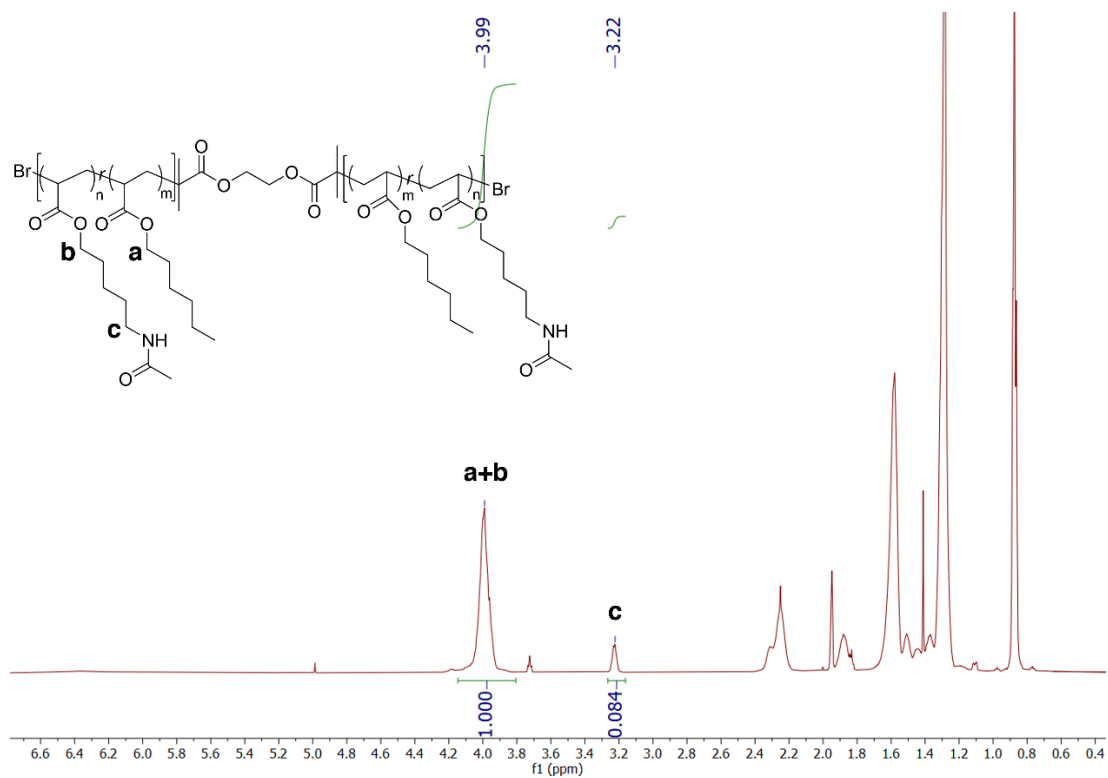

**Figure S9.** $^1$H NMR spectra of reversible middle block, HA$_{229}$-r-AAPA$_{21}$ in CDCl$_3$.

## 2.7 Gel permeation chromatography

We use gel permeation chromatography (GPC) to determine the polydispersity index (PDI) of polymers. For unentangle polymers, GPC measurements are performed in trifluoroethanol (TFE) with 0.02 M sodium trifluoroacetate (NATFAc) at 40°C using a TOSOH EcoSEC HLC-8320GPC system equipped with two isocratic pumps (one for the sample, the second for solvent) operated at 0.3 mL/min, a degasser, an auto sampler, one 4.6 mm × 35 mm TSKgel guard super AW-H column, two 6 mm × 150 mm TSKgel super AWM-H linear analytical columns with a 9 μm particle size, and a refractive index detector. The system is calibrated with poly(methyl methacrylate) standards.

For entangled polymer PHA, GPC measurement is performed using TOSOH EcoSEC HLC-8320GPC system with two TOSOH Bioscience TSKgel GMH$_{HR}$-M 5μm columns in series and a refractive index detector at 40°C. HPLC grade THF is used as the eluent with a flow rate of 1 mL/min. The samples are dissolved in THF with a concentration around 5 mg/mL.

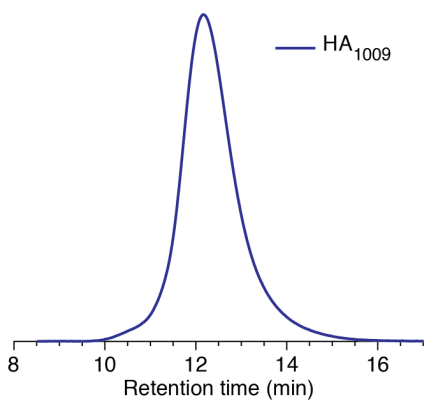

**Figure S10. GPC trace of entangled poly(hexyl acrylate) with DP of 1009.**

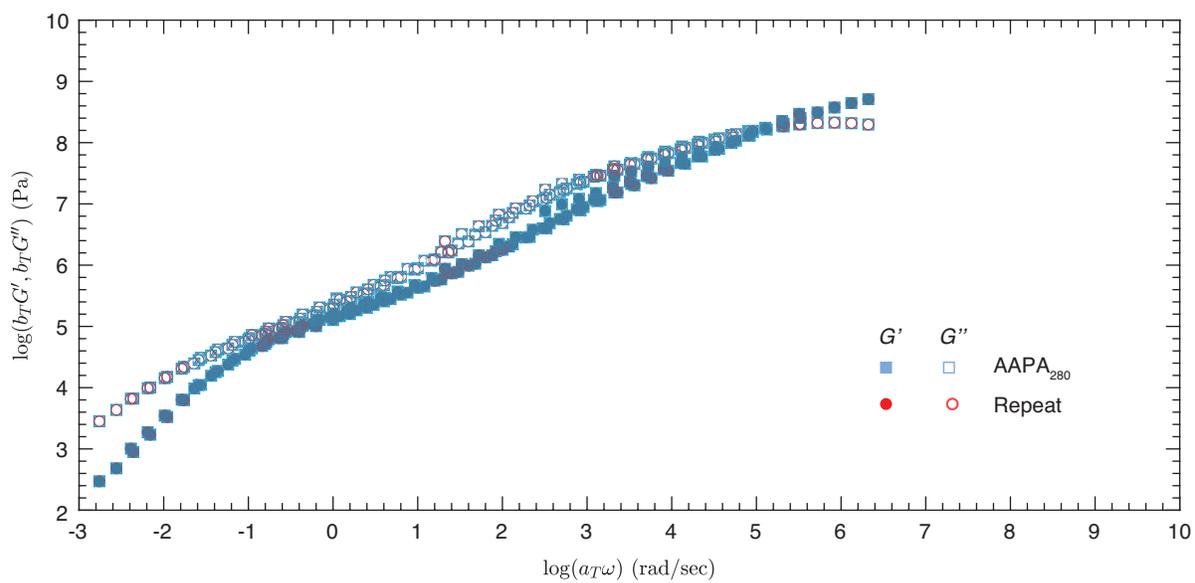

**Figure S11. Master curves of PAAPA with DP of 280 from two independent measurements.** The two curves perfectly overlap with each other.

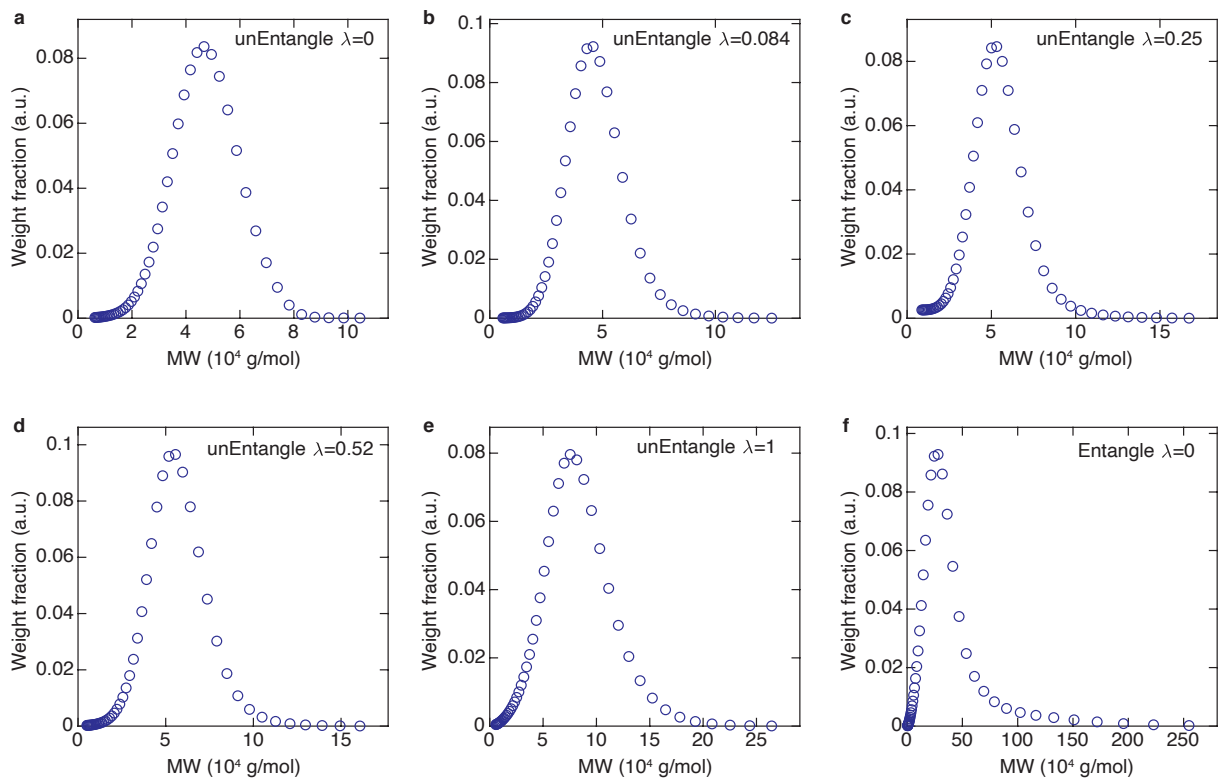

**Figure S12. Molecular weight distribution of unentangled and entangled polymers.**

# 3    ¹H NMR spectra of all polymers

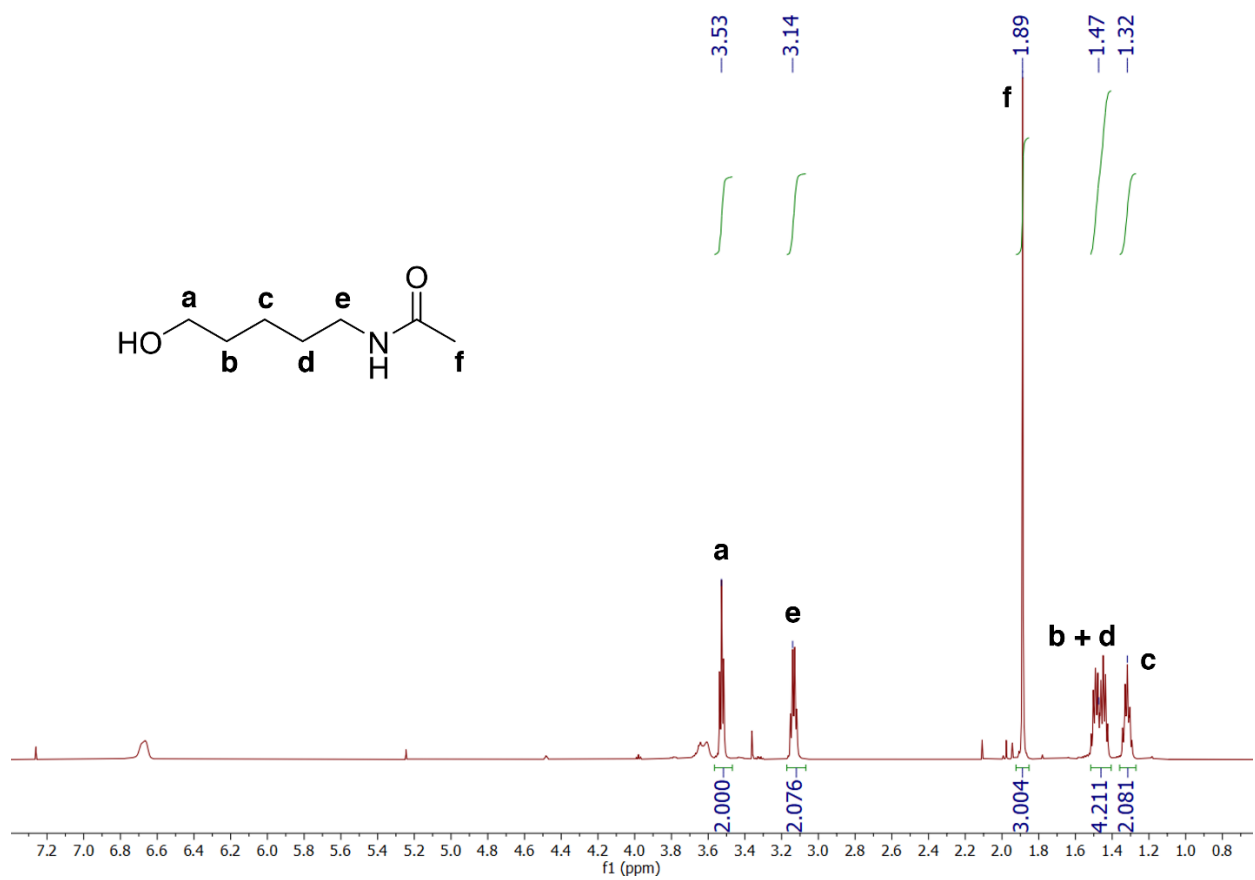

¹H NMR of 5-acetamino-1-pentanol in CDCl₃.

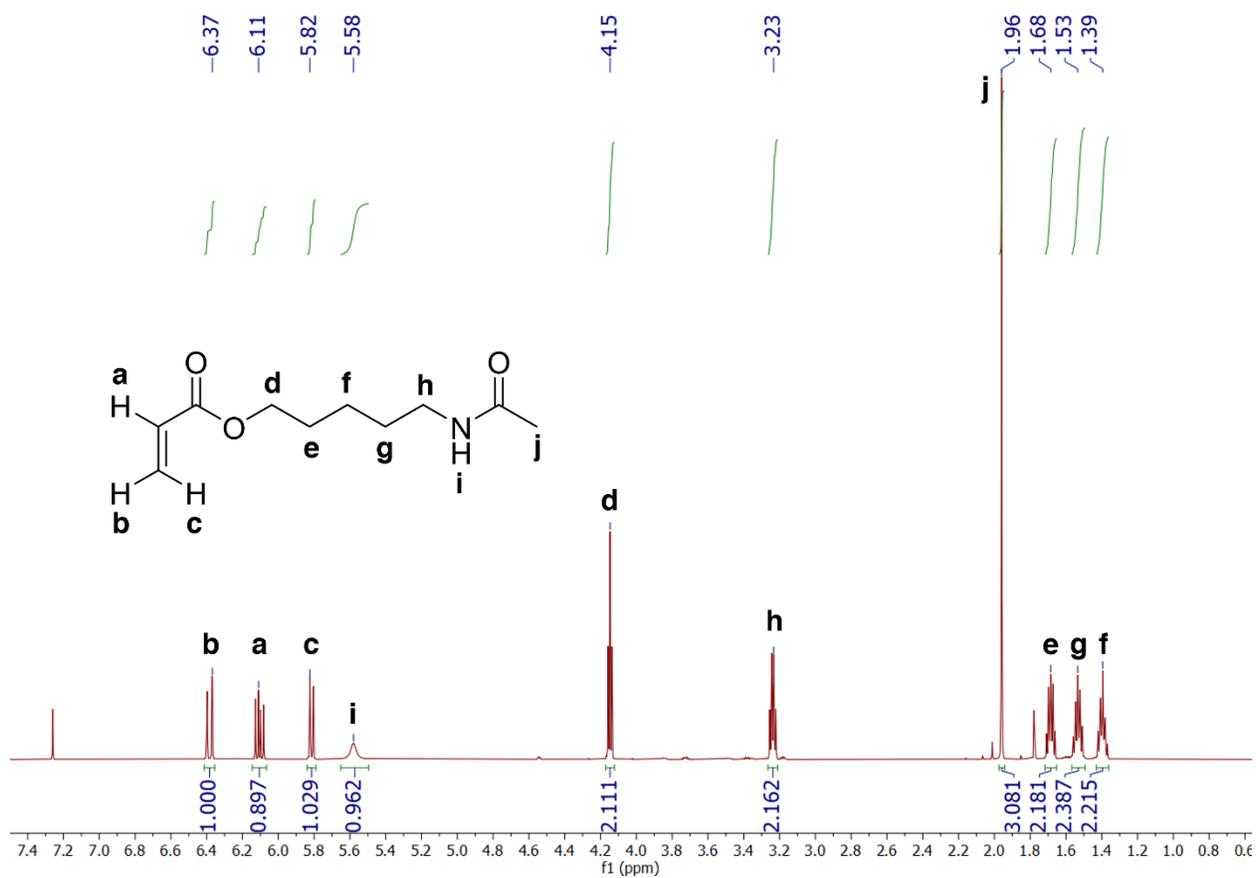

$^1$H NMR of 5-acetaminopentyl acrylate in CDCl$_3$.

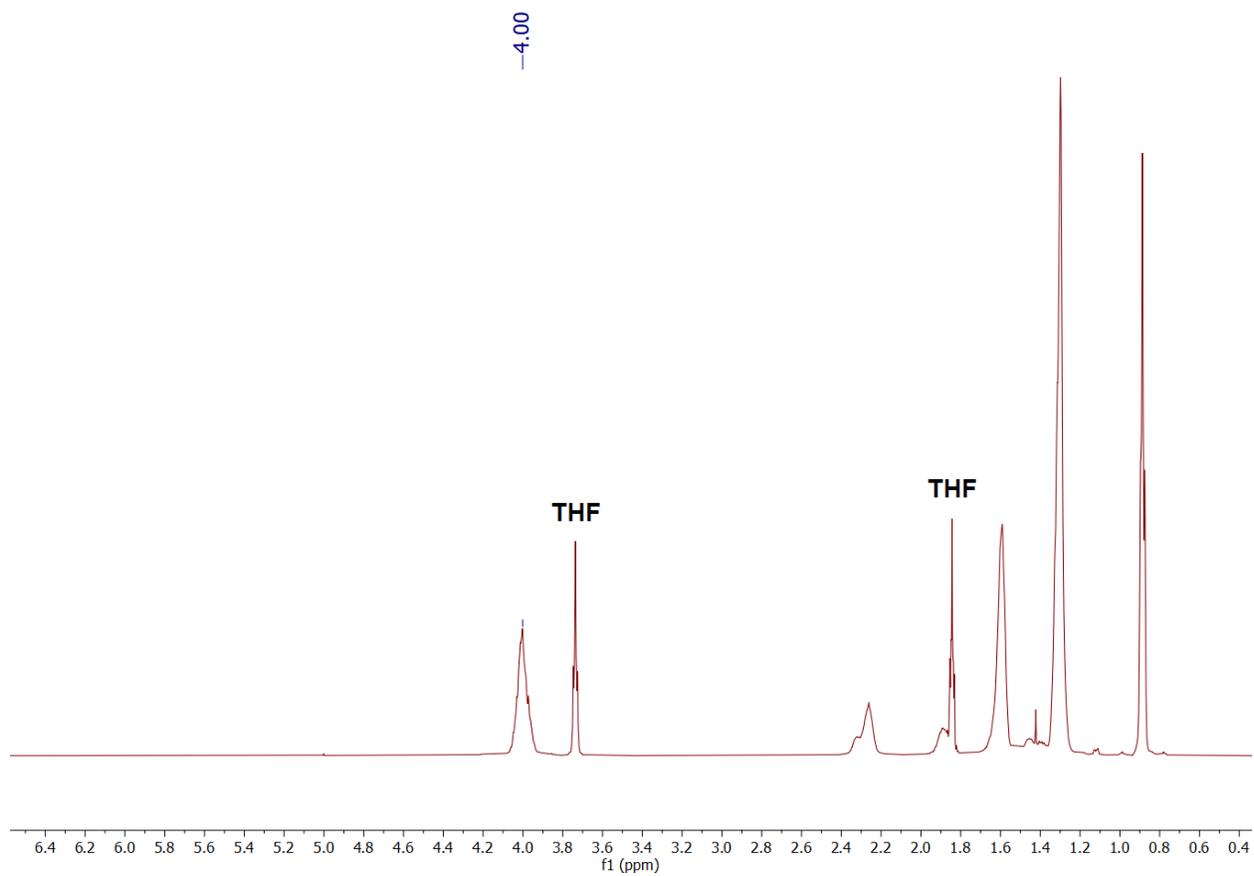

$^1$H NMR of HA$_{254}$ in CDCl$_3$.

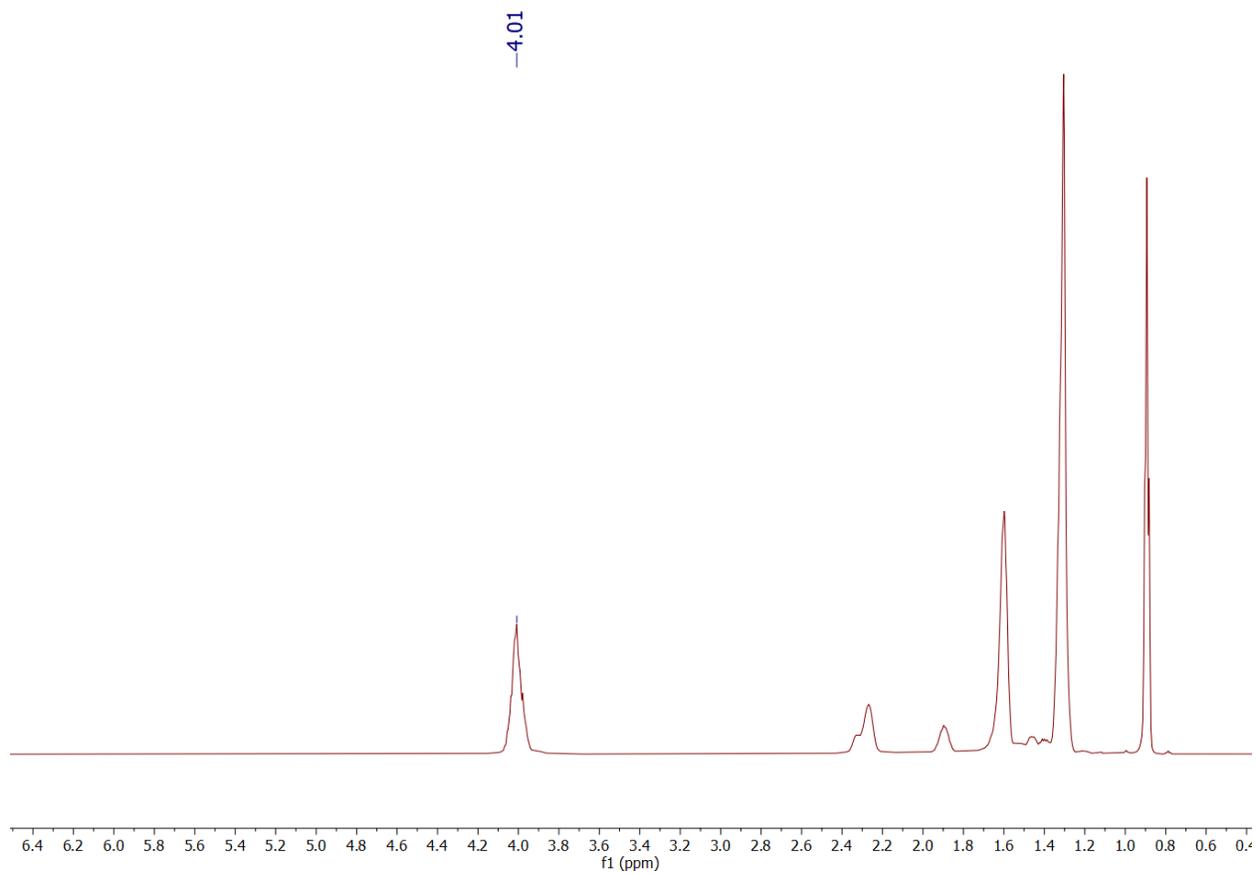

$^1$H NMR of HA$_{1009}$ in CDCl$_3$.

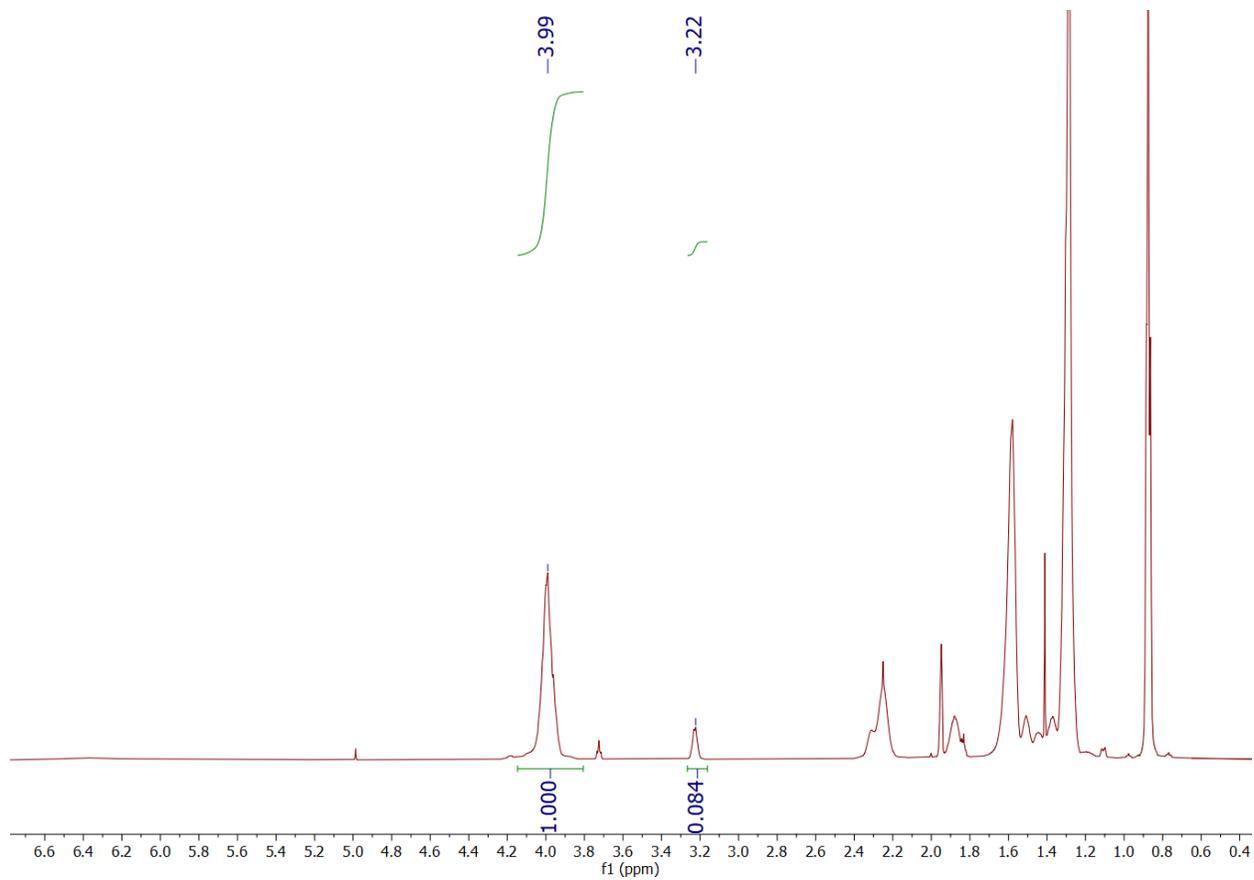

¹H NMR of HA₂₂₉-*r*-AAPA₂₁ in CDCl₃.

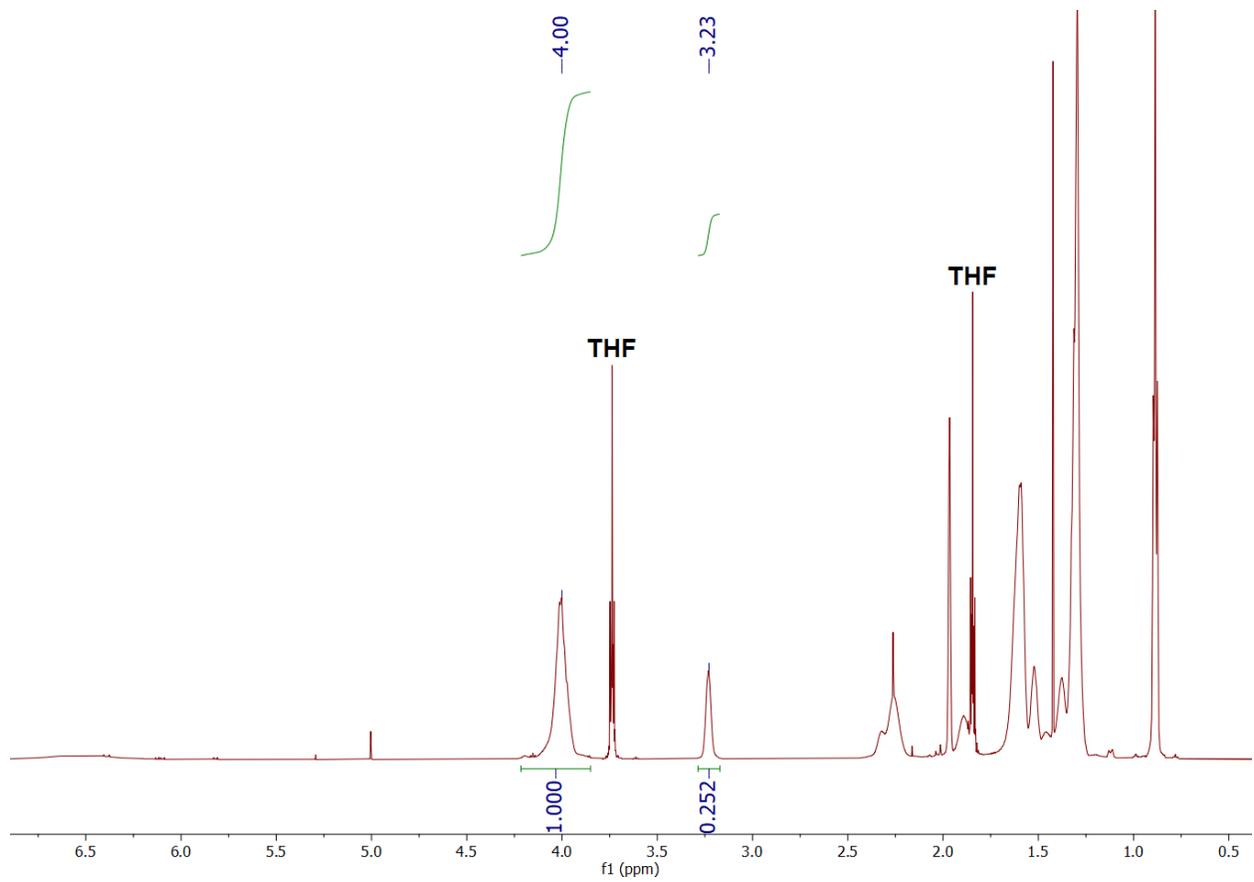

$^1$H NMR of HA$_{189}$-$r$-AAPA$_{64}$ in CDCl$_3$.

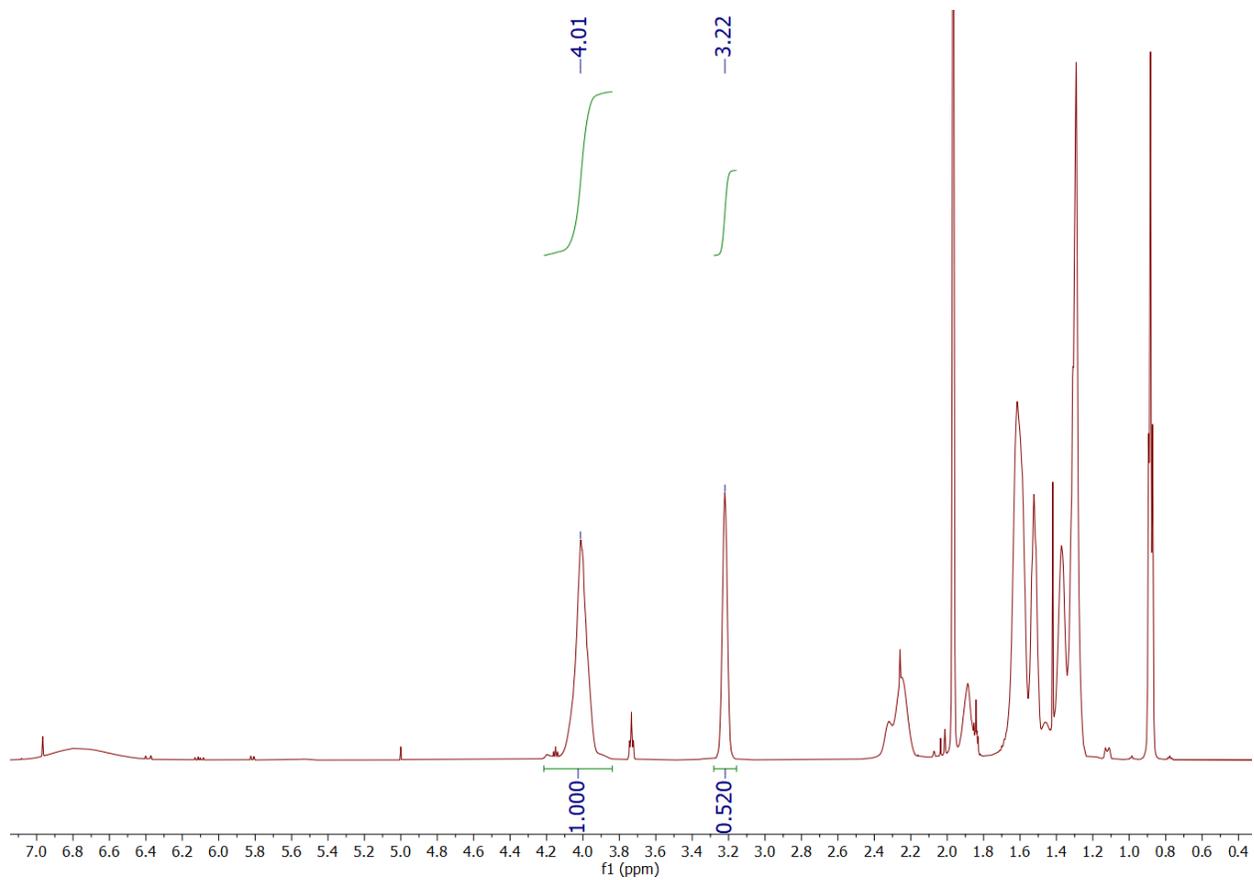

$^1$H NMR of HA$_{123}$-$r$-AAPA$_{133}$ in CDCl$_3$.

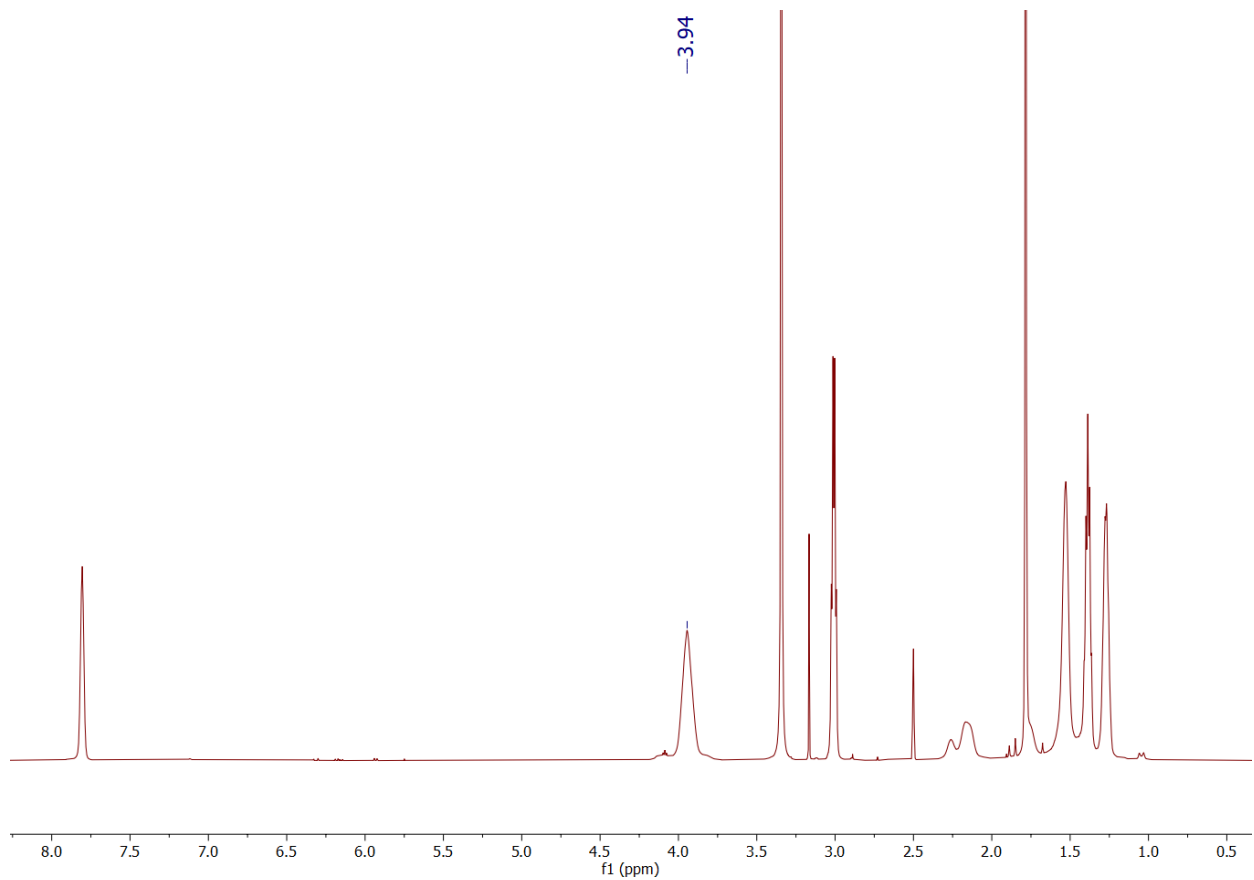

$^1$H NMR of AAPA$_{280}$ in DMSO-$d_6$.

**References**

[1] J. E. Mark, *Physical Properties of Polymers Handbook*, 2nd ed. (Springer Science & Business Media, 2007).

[2] M. Rubinstein and R. H. Colby, *Polymer Physics* (Oxford University Press, Oxford, UK, 2003).

[3] T. A. Kavassalis and J. Noolandi, *New View of Entanglements in Dense Polymer Systems*, Phys Rev Lett **59**, 2674 (1987).

[4] A. Afantitis, G. Melagraki, K. Makridima, A. Alexandridis, H. Sarimveis, and O. Iglessi-Markopoulou, *Prediction of High Weight Polymers Glass Transition Temperature Using RBF Neural Networks*, J Mol Struct THEOCHEM **716**, 193 (2005).

[5] T. G. Fox and P. J. Flory, *Second-Order Transition Temperatures and Related Properties of Polystyrene. I. Influence of Molecular Weight*, J Appl Phys **21**, 581 (1950).

[6] P. C. Hiemenz and T. P. Lodge, *Polymer Chemistry*, 2nd ed. (CRC Press, 2007).

[7] G. Williams and D. C. Watts, *Non-Symmetrical Dielectric Relaxation Behaviour Arising from a Simple Empirical Decay Function*, Trans Faraday Soc **66**, 80 (1970).

[8] R. Kohlrausch, *Theorie Des Elektrischen Rückstandes in Der Leidener Flasche*, Ann Phys **167**, 179 (1854).

[9] Y. Kawasaki, H. Watanabe, and T. Uneyama, *A Note for Kohlrausch-Williams-Watts Relaxation Function*, Nihon Reoroji Gakkaishi **39**, 127 (2011).

[10] W. L. Jorgensen, D. S. Maxwell, and J. Tirado-Rives, *Development and Testing of the OPLS All-Atom Force Field on Conformational Energetics and Properties of Organic Liquids*, J Am Chem Soc **118**, 11225 (1996).

[11] A. I. Jewett, D. Stelter, J. Lambert, S. M. Saladi, O. M. Roscioni, M. Ricci, L. Autin, M. Maritan, S. M. Bashusqeh, T. Keyes, R. T. Dame, J. E. Shea, G. J. Jensen, and D. S. Goodsell, *Moltemplate: A Tool for Coarse-Grained Modeling of Complex Biological Matter and Soft Condensed Matter Physics*, J Mol Biol **433**, 166841 (2021).

[12] A. P. Thompson, H. M. Aktulga, R. Berger, D. S. Bolintineanu, W. M. Brown, P. S. Crozier, P. J. in 't Veld, A. Kohlmeyer, S. G. Moore, T. D. Nguyen, R. Shan, M. J. Stevens, J. Tranchida, C. Trott, and S. J. Plimpton, *LAMMPS - a Flexible Simulation Tool for Particle-Based Materials Modeling at the Atomic, Meso, and Continuum Scales*, Comput Phys Commun **271**, 108171 (2022).

[13] K. Matyjaszewski, W. Jakubowski, K. Min, W. Tang, J. Huang, W. A. Braunecker, and N. V Tsarevsky, *Diminishing Catalyst Concentration in Atom Transfer Radical Polymerization with Reducing Agents*, Proc Natl Acad Sci U S A **103**, 15309 (2006).

[14] M. Zhernenkov, N. Canestrari, O. Chubar, and E. DiMasi, *Soft Matter Interfaces Beamline at NSLS-II: Geometrical Ray-Tracing vs. Wavefront Propagation Simulations*, in *Advances in Computational Methods for X-Ray Optics III*, International Society for Optics and Photonics, 9209, 92090G (2014).

[15] R. J. Pandolfi, D. B. Allan, E. Arenholz, L. Barroso-Luque, S. I. Campbell, T. A. Caswell, A. Blair, F. De Carlo, S. Fackler, A. P. Fournier, G. Freychet, M. Fukuto, D. Gürsoy, Z. Jiang, H. Krishnan, D. Kumar, R. J. Kline, R. Li, C. Liman, S. Marchesini, A. Mehta, A. T. N'Diaye, D. Y. Parkinson, H. Parks, L. A. Pellouchoud, T. Perciano, F. Ren, S. Sahoo, J. Strzalka, D. Sunday, C. J. Tassone, D. Ushizima, S. Venkatakrishnan, K. G. Yager, P. Zwart, J. A. Sethian, and A. Hexemer, *Xi-Cam: A Versatile Interface for Data Visualization and Analysis*, J Synchrotron Radiat **25**, 1261 (2018).